\documentclass[aps,preprint,superscriptaddress]{revtex4-1}
\usepackage{amsmath}
\usepackage{amssymb}
\usepackage{graphicx}
\usepackage{adjustbox}
\usepackage{soul}
\usepackage[normalem]{ulem}
\begin{document}
	
\title{Femtosecond laser produced periodic plasma in a colloidal crystal probed by XFEL radiation}
		
	\author{Nastasia Mukharamova}
	\affiliation{Deutsches Elektronen-Synchrotron DESY, Notkestra{\ss}e 85, D-22607 Hamburg, Germany}
		
	\author{Sergey Lazarev}
	\affiliation{Deutsches Elektronen-Synchrotron DESY, Notkestra{\ss}e 85, D-22607 Hamburg, Germany}
	\affiliation{National Research Tomsk Polytechnic University (TPU), pr. Lenina 30, 634050 Tomsk, Russia}
	
	\author{Janne-Mieke Meijer}
	\altaffiliation{Present address: Universiteit van Amsterdam, Science Park 904, 1090 GL Amsterdam, The Netherlands}
	\affiliation{Debye Institute for Nanomaterials Science, University of Utrecht, Padualaan 8, 3508 TB Utrecht, The Netherlands}

	\author{Oleg Yu. Gorobtsov}
	\altaffiliation{Present address: Cornell University, Ithaca, NY 14850, USA}
	\affiliation{Deutsches Elektronen-Synchrotron DESY, Notkestra{\ss}e 85, D-22607 Hamburg, Germany}

	\author{Andrej Singer}
	\altaffiliation{Present address: Cornell University, Ithaca, NY 14850, USA}
	\affiliation{University of California, 9500 Gilman Dr., La Jolla, San Diego, CA 92093, USA}

	\author{Matthieu Chollet}
	\affiliation{SLAC National Accelerator Laboratory, 2575 Sand Hill Rd, Menlo Park, CA 94025, USA}
		
	\author{Michael Bussmann}
	\affiliation{Institute of Radiation Physics, Helmholtz Zentrum Dresden-Rossendorf, 01328 Dresden, Germany}
    \affiliation{Center for Advanced Systems Understanding (CASUS), G\"{o}rlitz, Germany}

	\author{Dmitry Dzhigaev}
	\altaffiliation{Present address: Division of Synchrotron Radiation Research, Department of Physics, Lund University, S-22100 Lund, Sweden}
	\affiliation{Deutsches Elektronen-Synchrotron DESY, Notkestra{\ss}e 85, D-22607 Hamburg, Germany}

	\author{Yiping Feng}
	\affiliation{SLAC National Accelerator Laboratory, 2575 Sand Hill Rd, Menlo Park, CA 94025, USA}	
		
	\author{Marco Garten}
	\affiliation{Institute of Radiation Physics, Helmholtz Zentrum Dresden-Rossendorf, 01328 Dresden, Germany}
	\affiliation{Technische Universit\"{a}t Dresden, 01069 Dresden, Germany}
		
	\author{Axel Huebl}
	\altaffiliation{Present address: Lawrence Berkeley National Laboratory, 1 Cyclotron Rd, Berkeley, CA 94720, USA}
	\affiliation{Institute of Radiation Physics, Helmholtz Zentrum Dresden-Rossendorf, 01328 Dresden, Germany}
	\affiliation{Technische Universit\"{a}t Dresden, 01069 Dresden, Germany}

	\author{Thomas Kluge}
	\affiliation{Institute of Radiation Physics, Helmholtz Zentrum Dresden-Rossendorf, 01328 Dresden, Germany}
		
	\author{Ruslan P. Kurta}
	\altaffiliation{Present address: European XFEL, Holzkoppel 4, D-22869 Schenefeld, Germany}
	\affiliation{Deutsches Elektronen-Synchrotron DESY, Notkestra{\ss}e 85, D-22607 Hamburg, Germany}
	
	\author{Vladimir Lipp}
	\affiliation{Center for Free-Electron Laser Science, DESY, D-22607 Hamburg, Germany}
	
	\author{Robin Santra}
	\affiliation{Center for Free-Electron Laser Science, DESY, D-22607 Hamburg, Germany}
	\affiliation{Department of Physics, Universit\"{a}t Hamburg, 20355 Hamburg, Germany}

	\author{Marcin Sikorski}
    \altaffiliation{Present address: European XFEL, Holzkoppel 4, D-22869 Schenefeld, Germany}	
    \affiliation{SLAC National Accelerator Laboratory, 2575 Sand Hill Rd, Menlo Park, CA 94025, USA}

	\author{Sanghoon Song}
	\affiliation{SLAC National Accelerator Laboratory, 2575 Sand Hill Rd, Menlo Park, CA 94025, USA}

	\author{Garth Williams}
	\altaffiliation{Present address: NSLS-II, Brookhaven National Laboratory, Upton, NY 11973-5000, USA}
	\affiliation{SLAC National Accelerator Laboratory, 2575 Sand Hill Rd, Menlo Park, CA 94025, USA}
		
	\author{Diling Zhu}
	\affiliation{SLAC National Accelerator Laboratory, 2575 Sand Hill Rd, Menlo Park, CA 94025, USA}
	
	\author{Beata Ziaja-Motyka}
	\affiliation{Center for Free-Electron Laser Science, DESY, D-22607 Hamburg, Germany}
	\affiliation{ Institute of Nuclear Physics, PAS, Radzikowskiego 152, 31-342 Krakow, Poland}
	
	\author{Thomas Cowan}
	\affiliation{Institute of Radiation Physics, Helmholtz Zentrum Dresden-Rossendorf, 01328 Dresden, Germany}
	
	\author{Andrei V. Petukhov}
	\affiliation{Debye Institute for Nanomaterials Science, University of Utrecht, Padualaan 8, 3508 TB Utrecht, The Netherlands}
	\affiliation{Laboratory of Physical Chemistry, Department of Chemical Engineering and Chemistry, Eindhoven University of Technology P.O. Box 513, 5600 MB Eindhoven, Netherlands}
		
	\author{Ivan A. Vartanyants}
	\email[Corresponding author email: ]{ ivan.vartaniants@desy.de}
	\affiliation{Deutsches Elektronen-Synchrotron DESY, Notkestra{\ss}e 85, D-22607 Hamburg, Germany}
	\affiliation{National Research Nuclear University MEPhI (Moscow Engineering Physics Institute), Kashirskoe shosse 31, 115409 Moscow, Russia}

	\date{\today}
	
\begin{abstract}
	
With the rapid development of short-pulse intense laser sources, studies of matter under extreme irradiation conditions enter further unexplored regimes.
In addition, an application of X-ray Free-Electron Lasers (XFELs), delivering intense femtosecond X-ray pulses allows to investigate sample evolution in IR pump - X-ray probe experiments with an unprecedented time resolution.
Here we present the detailed study of periodic plasma created from the colloidal crystal.
Both experimental data and theory modeling show that the periodicity in the sample survives to a large extent the extreme excitation and shock wave propagation inside the colloidal crystal.
This feature enables probing the excited crystal, using the powerful Bragg peak analysis, in contrast to the conventional studies of dense plasma created from bulk samples for which probing with Bragg diffraction technique is not possible.
X-ray diffraction measurements of excited colloidal crystals may then lead towards a better understanding of matter phase transitions under extreme irradiation conditions.

\end{abstract}
	
\maketitle

\section{Introduction}

Studies of materials at high-pressure conditions above Mbar are highly relevant to the physics of shock compressed matter~\cite{ briggs2017ultrafast, stan2016liquid, kraus2016nanosecond, schropp2015imaging, koenig2005progress}, planetary formation~\cite{ross1981ice, coppari2013experimental, benedetti1999dissociation},  warm dense matter~\cite {fletcher2015ultrabright, kraus2016nanosecond, koenig2005progress}, and different types of plasma-matter interactions \cite{drake2006high, ciricosta2016measurements}.
The thermodynamic and transport properties of the high energy density material dictate its dynamics.
The general understanding of processes in materials under high pressure and temperature such as phase transitions~\cite{kraus2016nanosecond} or phase separations~\cite{kraus2017formation} are of a great scientific interest.
Theoretical investigations of the dynamics of materials under high pressure are widespread,
however, due to the limited diagnostic capabilities, experimental studies are still quite challenging.

There are currently  two major methods of generating extreme high pressure, that are the static compression with diamond anvil cells and dynamic (shock wave) compression.
The latter can be done for example by the powerful short-pulse lasers which offer the possibility of creating ultra-high pressure, much higher than achievable in static compression experiments.
The fundamental property of such high-power lasers is the creation of plasma at extreme pressure and temperature, which is causing a shock compression of the material.
Dynamic shock compression of  aluminum~\cite{fletcher2015ultrabright}, copper~\cite{kraus2016dynamic}, carbons~\cite{kraus2016nanosecond} and  hydrocarbons~\cite{kraus2017formation},  as well as other materials driven by high-power lasers is a subject of recent  extensive studies.
The shock wave speed is in the range of several kilometers per second, therefore, a facility providing picosecond resolution is required for the \textit{in situ} measurements of the shock-induced dynamics.

Newly developed X-ray Free-Electron Lasers (XFELs)~\cite {ackermann2007operation, emma2010first, ishikawa2012compact, tschentscher2017photon} are especially well suited for time-resolved measurements of the ultrafast structural dynamics of laser-created plasma~\cite{kraus2017formation, kluge2018observation}.
XFELs provide extremely intense coherent femtosecond X-ray pulses, which are necessary to perform experiments with a time resolution that outperforms synchrotron sources by orders of magnitude~\cite{vinko2012creation, fletcher2015ultrabright}.
X-ray scattering at XFEL is a powerful tool for successful studies of the rapid changes in the material caused by a high-power infrared (IR) laser in both space and time~\cite{dronyak2012dynamics}.
Although the scattering signal from the uniform plasma is not very high, sufficient response can be achieved if the plasma is periodically modulated in space thus allowing the use of much stronger X-ray Bragg scattering~\cite {petukhov2003bragg} and imaging~\cite{sulyanova2015structural, shabalin2016revealing, gulden2010coherent, bosak2010high, van2011scanning} techniques.

Such a unique form of matter as periodic plasma~\cite{lehmann2017laser} can be created, for example, by the high-power laser interaction with the periodically ordered dielectric material.
Recently, the properties of periodic plasma have been studied theoretically~\cite{sakai2007properties, gildenburg2019grating} and experimentally~\cite{kuo2007enhancement, monchoce2014optically, leblanc2016ptychographic, leblanc2017spatial}.
One of the fascinating properties of laser-produced periodic plasma is the enhancement of the generated intensity in the cases of low order harmonic generation in comparison to a uniform plasma~\cite{ganeev2014third}.
Therefore, investigation of dynamics and properties of a periodic plasma is beneficial for the development of laser-based radiation sources.
In the present study a periodic plasma was formed by the IR laser interaction with colloidal crystals made of polystyrene.

Polystyrene, consisting of carbon and hydrogen, is an ideal model system for creating a plasma by the IR laser sources because these atoms have a relatively low ionisation threshold, accessible by high-power IR laser.
This material is also of high relevance to the biological community, since most of the biological samples consist of light atoms such as carbon, nitrogen, oxygen, and hydrogen~\cite{rudenko2017femtosecond,abbey2016x}.
In addition, hydrocarbons are one of the most common chemical species throughout the Universe~\cite{kraus2018high}.
A considerable amount of hydrocarbons compressed to 150~GPa exists inside giant planets, especially icy giants such as Neptune and Uranus~\cite{helled2010interior}.
Also, some extrasolar planets~\cite{madhusudhan2012possible} and  white dwarf stars~\cite{dufour2007white}, are built from  high-pressure carbon which  was recently extensively studied~\cite{kraus2013probing}.

Here, we present \textit{in situ} IR pump--X-ray probe diffraction experiment performed at XFEL on a periodic polystyrene colloidal crystals.
The periodicity of the sample allowed us to apply the Bragg peaks analysis and to observe dramatic ultrafast changes in the colloidal crystal sample.
The experiment was performed with the  IR laser intensity on the order of $10^{14}$~W/cm$^2$.
With such high intensities a confined hot periodic plasma was created and generated a shock wave that compressed the surrounding pristine material.
This shock wave reached pressures on the order of 100~GPa, triggering fast changes in the colloidal crystal structure.
We found a good correspondence between the characteristic times determined in the experiment and in simulations.
Below we discuss the details of the pump-probe experiment and theoretical modeling of the processes involved in our studies.
\textbf{}
\clearpage
\section{Results}

\subsection{Pump-probe experiment}

The pump-probe experiment was performed at the Linac Coherent Light Source (LCLS)~\cite{emma2010first}  X-ray Pump Probe (XPP) beamline~\cite{chollet2015x} (see~\cite{mukharamova2017probing} and Methods for experimental details).
The colloidal crystal films were prepared from polystyrene spheres with a diameter of 163$\pm$3~nm, using the vertical deposition method~\cite{meijer2015colloidal}.
The grown colloidal crystal films consisted of 30-40 layers of close-packed hexagonal planes.
The thickness of the film was slightly depending on the position on a film along the growth direction.
The experiments were executed with three different IR laser intensities which are provided in Table~\ref{Table}.
For each IR laser intensity a pump-probe experiment was performed with a time delay variation $\tau$ from -10~ps to 1000~ps with 25.25~ps time increment.
Additionally, for two higher IR laser intensities ($I_2$ and $I_3$) we also accomplished measurements from -10~ps to 48.5~ps with the 6.5~ps time increment.
Due to a sample degradation, measurements for each time delay were performed at a new position of the sample.

Typical single-shot diffraction patterns for three different time delays are shown in the insets in Fig.~\ref{Experiment}.
Bragg peak parameters corresponding to the most intense 110~reflections such as integrated intensity ($I$) and the peaks Full Width at Half Maximum (FWHM)  in the radial ($w_q$) and azimuthal ($w_{\varphi}$) directions were analyzed.
For three IR laser intensities variation of the Bragg peak intensity  $\Delta I(\tau)/I$, radial $\Delta w_q(\tau)/w_q$, and azimuthal $\Delta w_\varphi(\tau)/w_\varphi$ broadening of the Bragg peaks are shown in Fig.~\ref{Results_Intensity} as a function of time delay (see Methods for the details  of the diffraction data analysis).

The experimental measurements in Fig.~\ref{Results_Intensity} are depicted by black dots for 25.25~ps time increment and by blue dots for 6.5~ps time increment.
The decay of the relative Bragg peak intensity accompanied by the growth of the peaks size (FWHM) for all three measured IR laser intensities is well visible.
For two higher IR laser intensities, an additional fast drop of the relative intensity during the first picoseconds was also observed.
From Fig.~\ref{Results_Intensity} this drop of intensity can be estimated to be about 10$\%$ of the initial intensity during the first 6.5 picoseconds.
This intensity drop was not accompanied by significant changes in the radial or azimuthal peaks size.

In order to obtain the characteristic timescales, results of our measurements were fitted with exponential functions.
For the lower IR laser intensity, the Bragg peak integrated intensity was fitted with one exponential function.
For the two higher IR laser intensities, the Bragg peak intensity decay could not be fitted with a single exponential function due to the fast drop during first picoseconds.
Therefore, these data were fitted with the two exponential functions which took into account both short and long characteristic timescales.
For the radial and azimuthal peaks sizes (FWHM) fitting was performed by a single exponential function for all IR intensities.
The results of the fits are shown in Fig.~\ref{Results_Intensity} by red lines and  they provide a good agreement with the experimental data.
For all three IR laser intensities the exponential fit of Bragg peak parameters such as intensity and peaks size (FWHM) provided  about 300-400~ps characteristic timescales (see Fig.~\ref{Results_Intensity} and Table~\ref{Table}).
For two higher IR laser intensities the short timescale on the order of 5~ps  was also revealed by the analysis of the Bragg peak intensities.

\subsection{Physics of high-power laser interaction with matter}

To further analyze the obtained scattering results, we propose the following model of the IR laser-matter interaction.
At the first stage the incoming high-power IR laser pulse ionizes the top layer of colloidal particles, creating a confined plasma on the top of the colloidal crystal.
The processes of plasma creation and expansion are modeled taking into account the periodicity of the  colloidal crystal sample and the polystyrene properties and will be discussed below in this section.

Polystyrene is a dielectric material and has no free electrons in the ground state.
It is also known to be transparent for the incoming IR pulses with 1.55 eV energy at low laser intensities.
However, the situation changes dramatically at high IR laser intensities.
The tight focusing of a 50~fs IR laser pulse with the energy about millijoules produces an IR laser intensity on the order of hundreds of terawatts per cm$^2$.
At such high IR laser intensities the so-called field ionisation is important.
It causes the plasma formation in the top layer of the colloidal crystal which may be described by the Keldysh theory~\cite{keldysh1965ionization}.
An important parameter of the theory is the so-called Keldysh parameter $\gamma$,
\begin{equation}
	\gamma=\frac{\omega_{IR} \sqrt{2 m_e E_{i}}}{E}=
	\begin{cases}
	> 1 \rightarrow \text {multi-photon regime}, \\
	< 1 \rightarrow \text{quasi-static regime}, 	
	\end{cases}
	\label{Eq:Keldysh}
\end{equation}
where  $\omega_{IR}$ is the frequency of the laser field, $m_e$ is the electron mass, $E_{i}$ is the zero-field ionisation energy of an atom, and $E$ is the electric field generated by the laser.
For low fields and high frequencies ($\gamma>1$) the multi-photon ionisation occurs, while for strong fields and low frequencies ($\gamma<1$) tunneling ionisation prevails.
Dependence of the Keldysh parameter on the IR laser intensity is shown in the Appendix Fig.~S3.
For all three IR laser intensities  and low ionisation states of carbon and hydrogen the Keldysh parameter is lower than unity (see Fig.~S3), so the quasi-static ionisation regime prevails~\cite{reiss2014tunnelling, reiss2008limits}.
The high-intensity IR laser field ionizes atoms in the polystyrene colloidal crystal up to H$^+$ and C$^+$ for the lower IR intensity and up to C$^{2+}$ for the two higher IR laser intensities (see Appendix).
As a result during first femtoseconds of the IR laser pulse propagation plasma is created on top of the colloidal crystal.

Due to the plasma creation, the incoming IR radiation is not penetrating any more into the colloidal crystal sample, due to the well-known plasma skin effect~\cite{kittel2005introduction, pitaevskii2012physical}.
The depth of the skin layer $l_{skin}$ depends on  the frequency of the IR laser $\omega_{IR}$ and plasma frequency $\omega_p=\sqrt{n_e e^2/m_e \varepsilon_{0}}$, where $n_e$ is the number density of electrons,  $e$ is the electric charge, and  $\varepsilon _{0}$ is the permittivity of free space.
In our case, assuming single ionisation of each atom, the skin depth $l_{skin}= c/\sqrt{(\omega_p^2-\omega_{IR}^2)}$ is about 10-20~nm.	
Free electrons formed in the skin layer by the strong laser-matter ionisation process are further accelerated by the inverse bremsstrahlung~\cite{krainov2002cluster} and resonance absorption~\cite{gibbon1996short} mechanisms.
As such,  high-energetic electrons are propagating inside the first layer of the colloidal particles of the crystal.		
Accelerated electrons collide inelastically with the atomic ions inside the  colloidal particle which causes additional collisional ionisation of the C atoms up to C$^{4+}$ (see Appendix for further details).
Ionisation and IR laser energy absorption processes described above occur within the colloidal crystal which has a periodic structure.
As a result, created plasma also has the same periodicity as the colloidal crystal during the first picoseconds after the IR laser pulse interaction with a colloidal crystal.

In order to simulate the first stages of creation and dynamics of the periodic plasma we used the Particle-In-Cell on Graphic Processor Units (PIConGPU) code version 0.4.0-dev developed at Helmholtz-Zentrum Dresden-Rossendorf (HZDR)~\cite{PIConGPU2013, burau2010picongpu}.
PIConGPU simulations were performed in the time interval from 0 to 1~ps for all three IR laser intensities measured in our pump-probe experiment.
Two different types of ionisation processes are dominating in the colloidal crystal, namely field ionisation and collisional ionisation, and they were included in the simulations.
The ionisation rate of the field ionisation process was calculated according to the Ammosov-Delone-Krainov (ADK) model~\cite{delone1998tunneling} and the collisional ionisation was simulated using the Thomas-Fermi ionisation model~\cite{more1985pressure} (see Appendix for details).

The electron energy density distribution as a function of depth and transverse spatial coordinate at 80~fs and 1~ps after the start of the IR laser pulse propagation are shown in Fig.~\ref{plasma}.
As one can see from Fig.~\ref{plasma}(a-c), at 80~fs only the first layer of the colloidal particles is strongly ionized by the IR laser pulse.
As a result, the periodic plasma is formed on top of the colloidal crystal.
Our simulations show that the maximum electron energy density was reached in the center of colloidal particles of the first layer 80~fs after the start of the IR laser pulse propagation (see Fig.~\ref{plasma}(a-c)).
The maximum electron energy density is summarized in Table~\ref{Table} for three measured IR laser intensities.
The pressure reaches its maximum in the first layer in the center of each colloidal particle, thus forming a periodic plasma state.

At 1 ps after the beginning of the interaction of the IR laser with the colloidal crystal sample, accelerated electrons move deep inside the colloidal crystal and collisionally ionize the inner part of the crystal (see Fig.~\ref{plasma}(d-f) and Appendix for further details).
The electron energy density at 1 ps has its maximum still in the first layer of the colloidal crystal but its magnitude is much lower than at 80~fs (see Table~\ref{Table}).
Even after 1~ps the electron energy density distribution resembles a periodic structure of the colloidal crystal.

To determine the evolution of the electron energy density distribution, we averaged simulated values over the transverse x- and y-coordinates.
The time dependence of the maximum electron energy density is shown in Fig.~\ref{Pressure_time_z}(a).
As shown in this figure the electron energy density reaches its maximum value at 80~fs for all IR laser intensities values, and after 0.6~ps remains practically constant.
The z-dependence at 1~ps is shown in Fig.~\ref{Pressure_time_z}(b).
It is clearly seen that the electron energy density is decaying along the z-direction but periodic modulations due to the colloidal crystal structure are well visible.
This periodicity is less visible for the lower IR laser intensity because of the low ionisation level of the inner part of the colloidal crystal.

\clearpage
\subsection{Ablation and shock wave propagation}

After the plasma formation, the electron-ion thermalisation occurs and further dynamics in the colloidal crystal is governed by the hydrodynamics.
These processes are ablation of the material and shock wave propagation inside the cold material.
This last hydrodynamic stage of our model, which was also observed in our experiment, will be discussed in this section.

As can be seen from Fig.~\ref{Pressure_time_z}(a) the ionisation of the colloidal crystal was practically finished at 1~ps.
Around these times the high-pressure dense plasma in the top layers of the colloidal crystal induced ablation and shock wave propagation inside the sample.
As a result, the top layers of the sample were ablated and a strong shock wave compressed the solid and destroyed the periodicity of the inner part of the colloidal crystal.
We relate experimentally observed fast drop of the scattered intensity to the ablation process of the top layers of the colloidal crystal.
The shock wave propagation manifests itself as an exponential drop of intensity with the typical time scales on the order of hundreds of picoseconds (see Fig.~\ref{Results_Intensity}) in the IR pump--X-ray probe diffraction experiment.

In order to model  structural changes in the colloidal crystal, hydrodynamic simulations using the HELIOS code were performed~\cite{macfarlane2006helios} (see Methods for simulation details).
The HELIOS code is widely used to simulate the dynamics of plasma evolution created in high-energy density physics experiments~\cite{kraus2017formation,kraus2018high,kraus2016nanosecond}.
The hydrodynamic simulation was one-dimensional, therefore the 3D structure of the colloidal crystal was modeled as layers with the periodic variation of a mass density.
The simulations were performed using two-temperature model, which takes into account the fact that the energy of hot electrons is not instantaneously transferred to cold ions, but is governed by the electron-phonon coupling.
The pressure and mass density evolution obtained from the hydrodynamic simulations are shown in Fig.~\ref{HELIOS} (see also Appendix).

The first process that was determined in the hydrodynamic simulations is the ablation of the material on the top of the colloidal crystal.
The ablation threshold of polystyrene for 800~nm laser wavelength with 40~fs pulse duration as reported in \cite{suriano2011femtosecond} is on the order of 10 mJ/cm$^2$.
The laser fluences used in our pump-probe experiment were three orders of magnitude higher than the polystyrene ablation threshold (see Table~\ref{Table}).
During the first picoseconds the top layers of the colloidal crystal were already damaged by the ablation process, and we observe a steep gradient of the mass density in our hydrodynamic simulations (see in Fig.~\ref{HELIOS}(d-f), and Appendix for details).
After the first picoseconds the ablation process continues and results in a zero mass density on the top of the sample.
Due to the ablation process of the top layers of the colloidal crystal we observed a fast initial drop of the scattered intensity in Fig.~\ref{Results_Intensity}(b,c).
The ablation process stops at about 180 nm -- 450 nm depth which corresponds to about 1-3 layers (see Table~\ref{Table}).

The next process that occurs is the shock wave propagation inside the periodic colloidal crystal.
As one can see from Fig.~\ref{HELIOS}(a-c), the shock pressure is propagating along the z-direction and destroys the periodicity of the significant part of the sample.
During the first picoseconds the maximum shock wave pressure is located in the top layer of the colloidal crystal (see Fig.~\ref{Pressure_time_z},~\ref{HELIOS}(a-c)).
The shock wave speed is proportional to the square root of pressure, and it was about 6~km/s on the top of the sample and about 4~km/s on the border of the shock wave front with the cold material.
Therefore, around 100~ps at the depth of about one micrometer (that corresponds to about 8~layers) the high pressure front reaches the low pressure front (see Fig.~\ref{HELIOS} and Appendix).
After 100~ps the high pressure front propagates further inside the sample and destroys the sample periodicity.
The average shock wave propagation speed obtained from our simulations is on the order of 5~km/s and maximum mass velocity is on the order of 2~km/s which is in a good agreement with previous studies~\cite{marsh1980lasl}.

While  propagating inside the colloidal crystal, the shock wave is losing its speed due to dissipation of energy and consequently the pressure of the shock wave front is decreasing gradually.
At the distance where the shock wave pressure is not sufficient to compress the colloidal crystal, the shock wave effectively stops.
For three IR laser intensities used in this experiment, the shock wave stopped after approximately 400~ps -- 900~ps propagation time at a depth of 2~$\mu$m -- 5~$\mu$m  and the exact values are provided in Table~\ref{Table}.
The sample periodicity was not further destroyed beyond this point, because the shock wave converts into a sound wave, which does not induce any structural transformation of the colloidal crystal sample~\cite{gamaly2013physics}.

In order to study the influence of the sample periodicity on the shock wave propagation, we performed hydrodynamic simulations for the non-periodic polystyrene sample.
The results were obtained for all three IR laser intensities used in our XFEL experiment and are summarized in Appendix.
From the comparison of these two sets of simulations we can conclude that for the non-periodic sample the shock wave stops earlier and the depth of the shock wave propagation is about 30\% smaller than for the periodic one.
The ablation depth is also about twice shorter for the non-periodic sample (see Appendix Fig.~S15-S17 and Table~SIV).
Such difference can be explained by the higher average mass density in the case of the non-periodic sample.
A comparison of two simulation sets shows that the pressure modulations observed in Fig.~\ref{HELIOS}(a-c) are caused by the periodic structure of the colloidal crystal sample.



\section{Discussion}

In the present work we studied experimentally and by theoretical modelling the dynamics of the periodic plasma induced by high-power IR laser in the polystyrene colloidal crystals.
We performed a pump-probe diffraction experiment at LCLS on a periodic plasma created from a colloidal crystal sample.
The periodic structure of the colloidal crystal allowed us to measure Bragg peaks from the sample.
We observed a fast decay of the Bragg peak intensity and the growth of the radial and azimuthal peaks width.
Such changes of scattering parameters indicate ultrafast dynamics of the colloidal crystal periodic structure, which is producing the scattering signal.
From the analysis of the Bragg peak parameters we obtained  5~ps short and 300~ps long characteristic timescales.

\begin{figure}
\includegraphics[width=\linewidth]{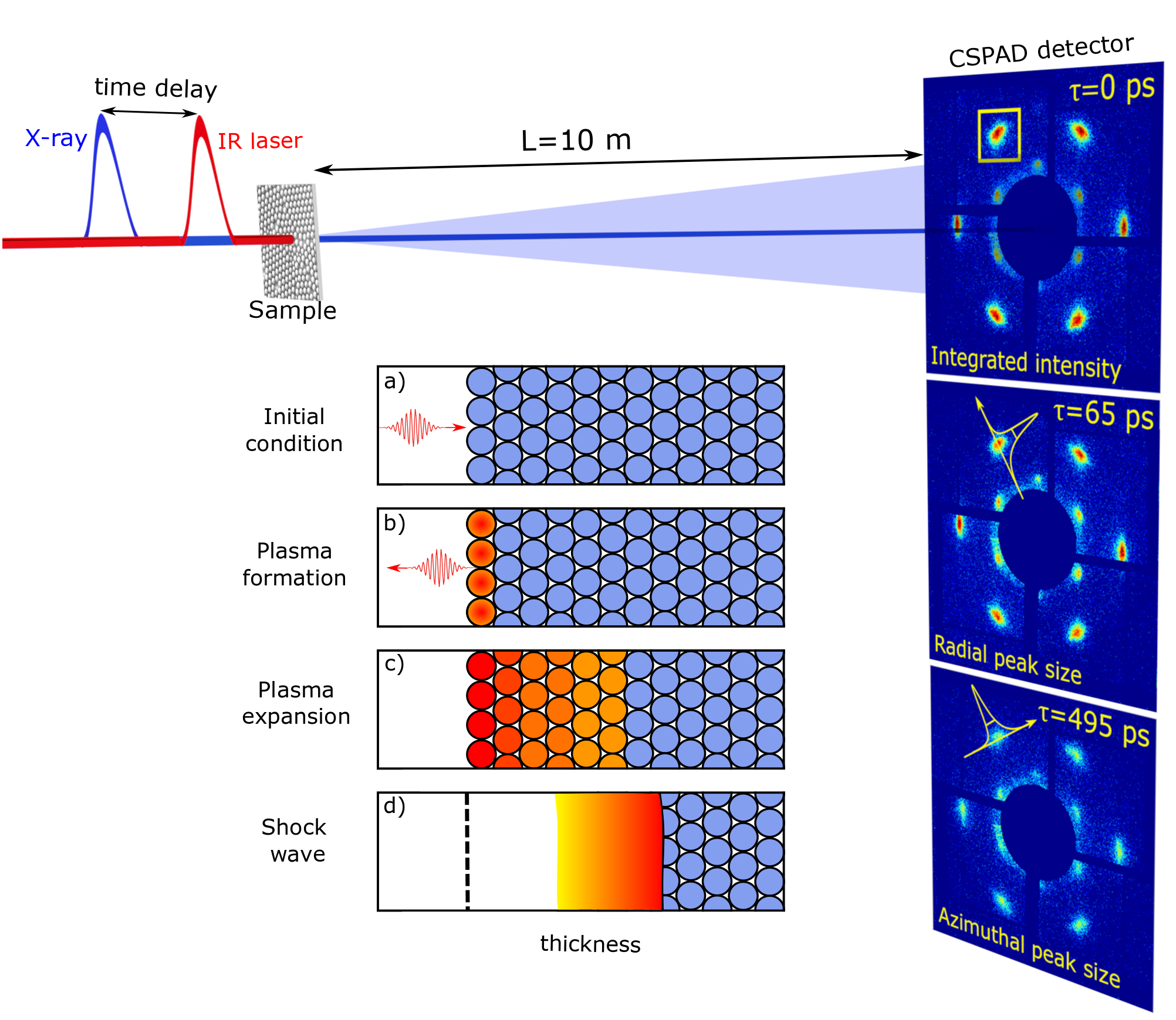}
\caption{
Scheme of the pump-probe experiment.
XFEL pulses generated by the undulator are monochromatized by the diamond crystals and focused by the compound refractive lenses (not shown) to the size of 50~$\mu$m at the sample position.
CSPAD detector is positioned 10~m downstream from the colloidal sample.
Evolution of diffraction patterns as a function of time delay between the IR pump laser and X-ray Probe laser is shown on the right.
Insets (a-d).
Three-stage model of the IR laser-matter interaction.
The colloidal particles are shown as circles.
The color of the particles corresponds to the temperature of the colloidal crystal - red is plasma and blue is the cold material.
The incoming IR laser pulse is pointing in the direction of the pulse propagation.
Initially, the IR laser pulse is propagating towards the colloidal crystal sample and after interaction with the sample it is reflected by the created plasma on the top layer of the colloidal crystal.
The top surface level of the initial colloidal crystal is marked by the black dashed line in (d).	
}
\label{Experiment}
\end{figure}

\begin{figure}
\includegraphics[width=\linewidth]{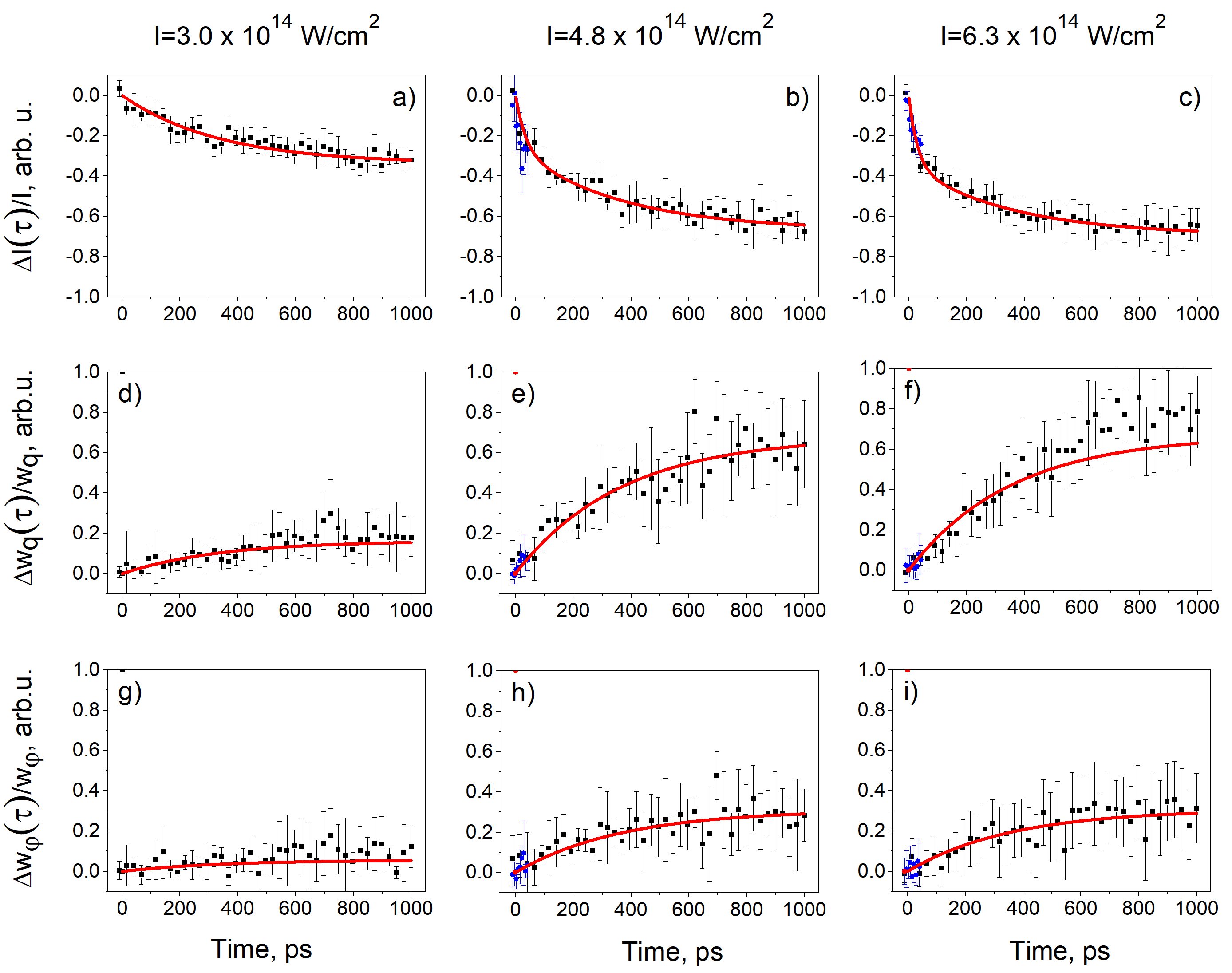}
\caption{Time dependence of the relative change of the integrated intensity of the Bragg peaks $\Delta I(\tau)/I$ (a-c) and their widths in the radial $\Delta w_q(\tau)/w_q$ (d-f) and azimuthal $\Delta w_{\varphi}(\tau)/w_{\varphi}$ (g-i) directions at three measured IR laser intensities.
Black (blue) dots  are experimental data corresponding to 25.25~ps (6.5~ps) time delay increment and solid red lines are exponential fits.
}
\label{Results_Intensity}
\end{figure}

\begin{figure}
\includegraphics[width=\linewidth]{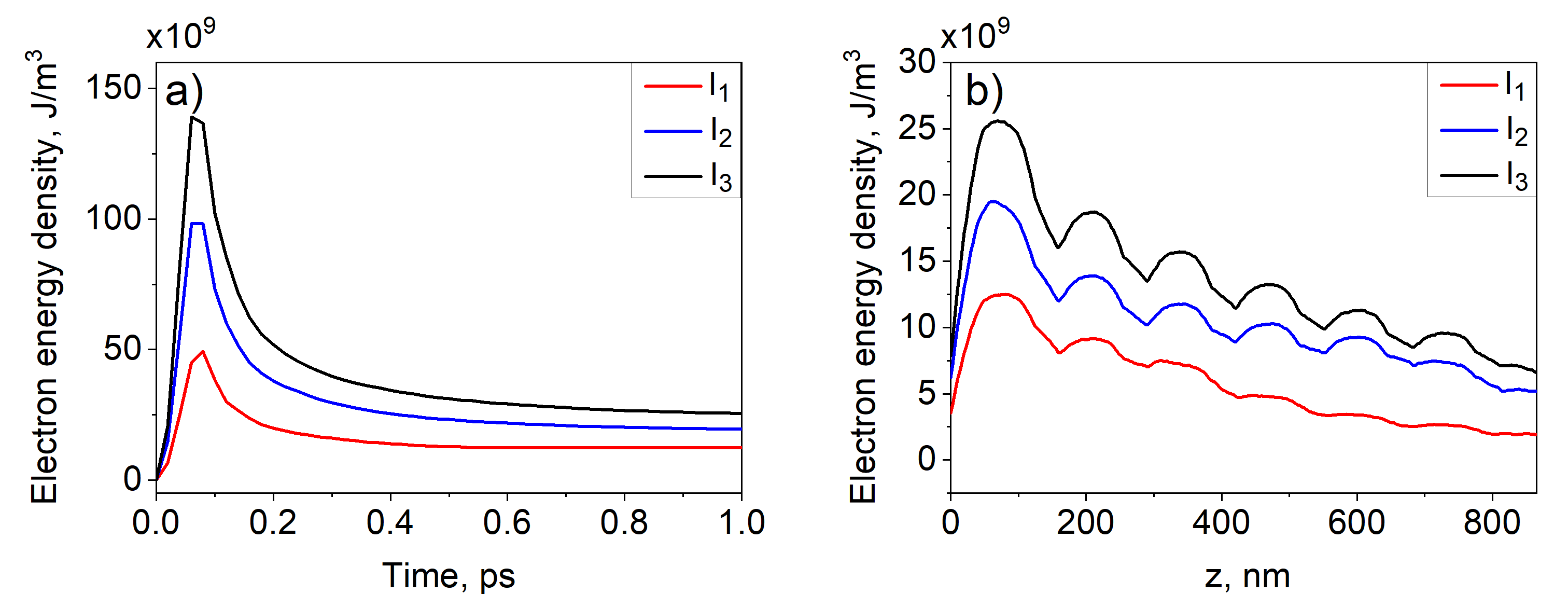}
\caption{Electron energy density distribution in the colloidal crystal at 80~fs (a-c) and 1~ps (d-f) after the start of the IR laser pulse for three different IR laser intensities.
The IR laser pulse is coming from the top along the z direction.
Here we show a projection of the electron energy density along the y-direction.
}			
\label{plasma}
\end{figure}

\begin{figure}
\includegraphics[width=\linewidth]{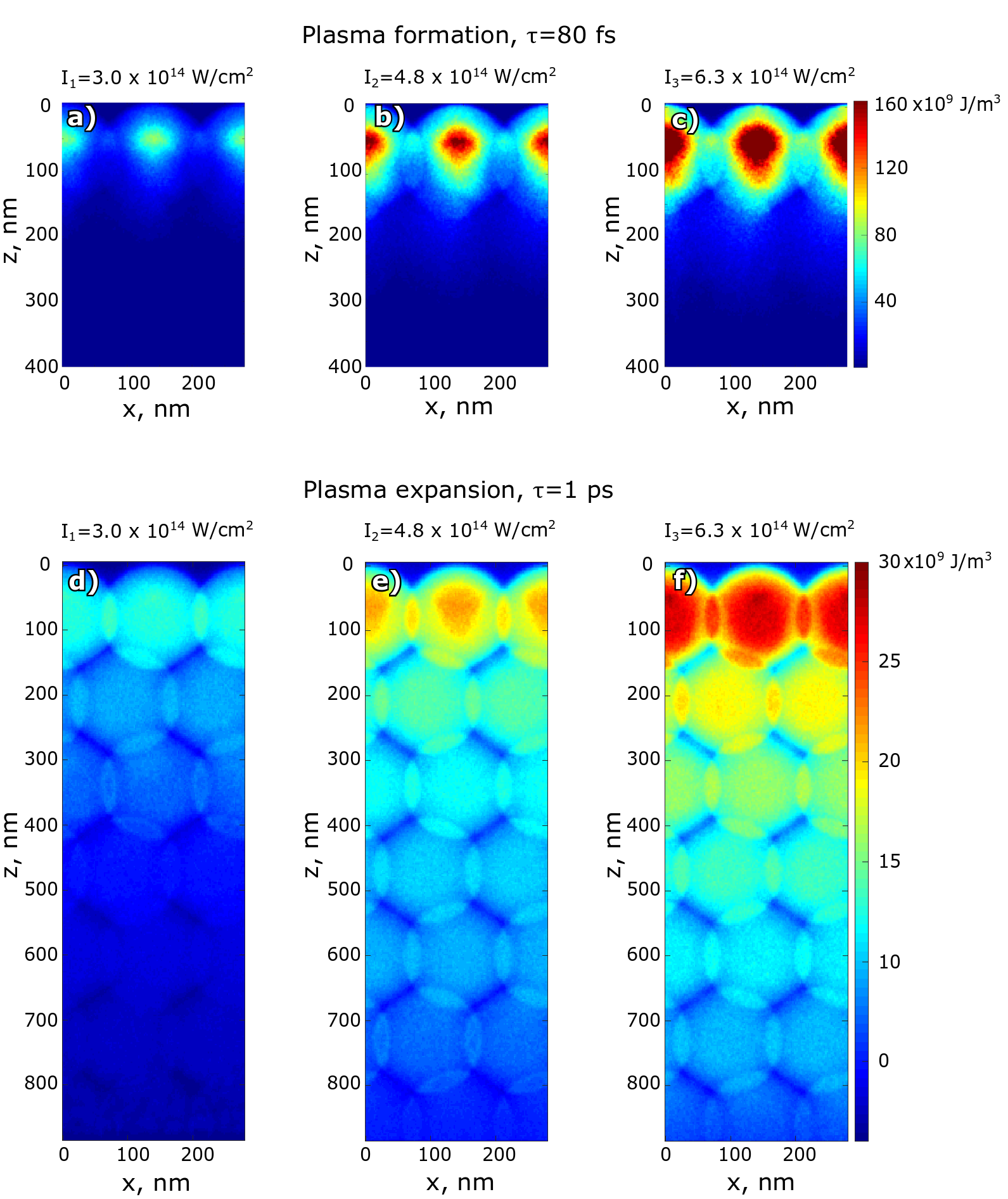}
\caption{Time (a) and depth (b) dependencies of the electron energy density  for three measured intensities $I_1=3.0 \cdot 10^{14}$ W/cm$^2$, $I_2=4.8 \cdot 10^{14}$ W/cm$^2$ and $I_3=6.3 \cdot 10^{14}$ W/cm$^2$.
Time dependencies of the electron pressure are shown at 75 nm distance from the top of the sample that correspond to the center of colloidal particles in the first surface layer.
Electron energy density-depth dependence is shown at 1 ps after start of the interaction with the IR laser pulse.
}	
\label{Pressure_time_z}
\end{figure}

\begin{figure}
\includegraphics[width=\linewidth]{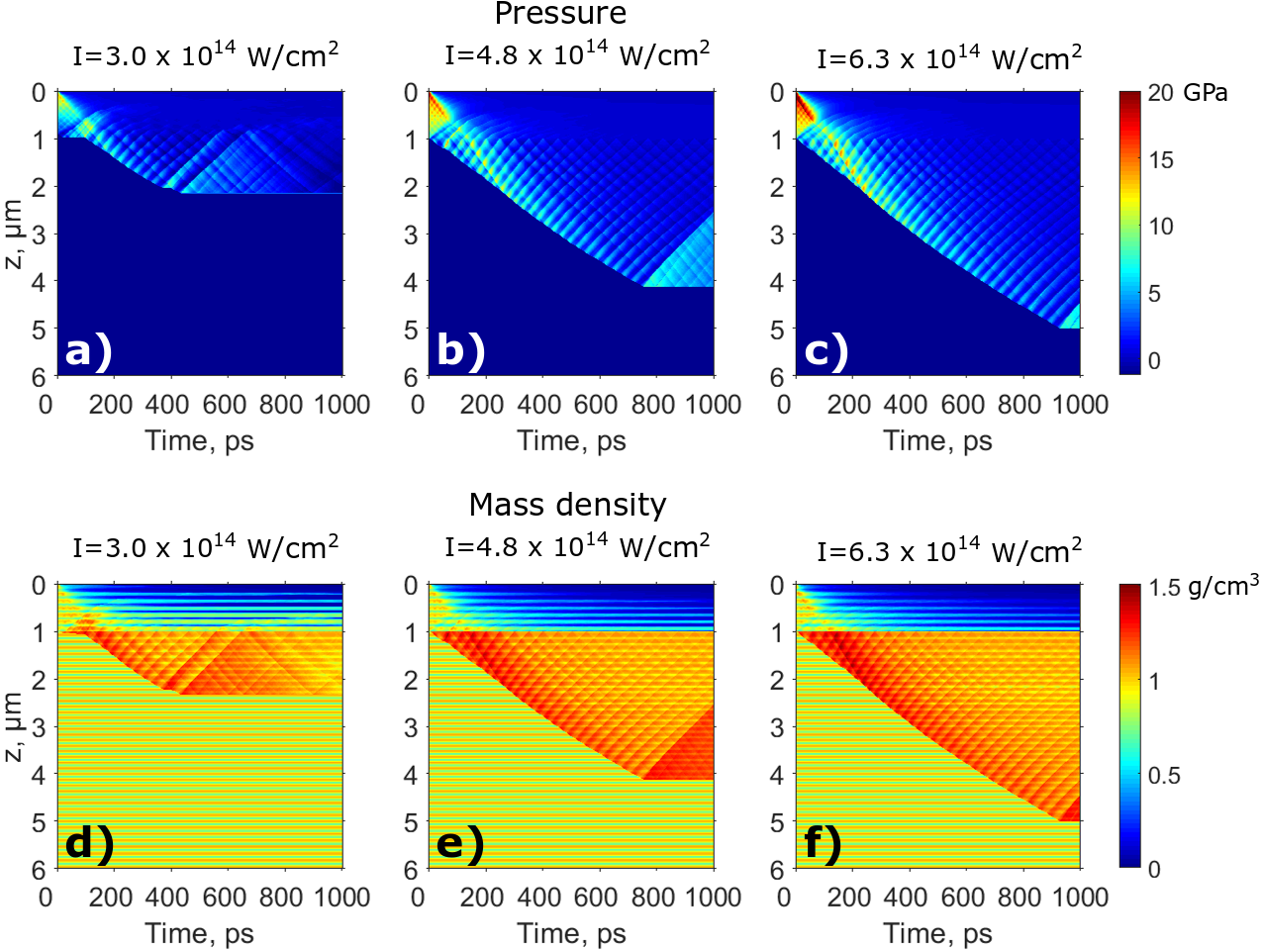}
\caption{Hydrodynamic simulations of the shock wave propagation. Color plots show simulation results for the pressure  (a-c) and mass density (d-f) for three different IR laser intensities: (a,d)~$I_1$=$3.0 \cdot 10^{14}$ W/cm$^2$,  (b,e)~$I_2$=$4.8 \cdot 10^{14}$ W/cm$^2$, (c,f)~$I_3$=$6.3 \cdot 10^{14}$ W/cm$^2$.   }			
\label{HELIOS}			
\end{figure}

We have proposed a three-stage model of interaction between the high-power IR laser pulse and the periodic colloidal crystal to explain the ultrafast changes in the colloidal sample (see Fig.~\ref{Experiment}).
First, all colloidal particles are in the initial condition, unaffected by the laser pulse.
The incoming IR laser pulse generates a plasma in the top layer of the colloidal crystal within the first few femtoseconds (see Fig.~\ref{Experiment}(a,b)).
Due to the plasma skin effect the IR laser pulse is partially reflected (see Fig.~\ref{Experiment}(b)).
This hot confined periodic plasma then propagates inside the colloidal crystal up to about 0.6~ps time (see Fig.~\ref{Experiment}(b)).
The second stage is the ablation of a few layers on the top of the colloidal crystal.
The top layers of the sample are damaged during the first few picoseconds and are completely destroyed afterwards (see Fig.~\ref{Experiment}(d)).
During the third and last stage the shock wave is formed and propagates inside the colloidal crystal sample.
This shock wave destroys the periodicity of the sample by compressing the structure in its deeper parts (see Fig.~\ref{Experiment}(d)).

The three-stage model of the laser-matter interaction allowed us to attribute the short 5~ps time scale determined in our diffraction pump-probe experiment to the ablation of the material.
The long 300~ps time scale is related to the shock wave propagation.
Simulations were performed for all three stages of the laser-matter interaction: plasma formation and expansion were simulated with the 3D PIConGPU code, ablation and shock wave propagation were simulated with the HELIOS code.
From the results of plasma and hydrodynamic simulations the time dependence of the structural changes in the colloidal crystal was obtained.
The results of simulations are in a good agreement with the analysis of the Bragg peaks, extracted from the diffraction patterns measured in our pump-probe experiment (see Fig.~\ref{Results_Intensity} and Table~\ref{Table})).

In the hydrodynamic simulations, the shock wave stops at different depth of the colloidal crystal depending on the incoming IR laser intensity (see Fig.~\ref{HELIOS} and Table~\ref{Table}).
As a result, the amount of colloidal crystal affected by the shock wave is higher for higher IR laser intensity and this is consistent with the experimental data.

Finally, we demonstrated that shock wave propagation inside the periodic colloidal crystal can be visualized \textit{in situ} with a high temporal resolution by an IR pump -- X-ray probe experiment at an XFEL facility.
The periodic structure of the colloidal crystal allowed us to reveal the picosecond dynamics of the propagating shock wave by Bragg peak analysis.
We obtained short and long characteristic timescales corresponding to the ablation of the material and shock wave propagation, respectively.
At the same time our simulations predict much shorter times of evolution of plasma and ablation processes in polystyrene colloidal samples.
This is still an open and intriguing question of investigation of the plasma dynamics and ablation process with sub-picosecond time resolution and will need special attention in future experiments.

By performing IR-pump and X-ray probe experiments on the periodic samples we foresee that formation and development of the periodic plasma may be studied in detail in future.
The application of these ideas and methodology based on scattering from the periodic samples may lead towards new ways of investigating of phase transitions in matter under extreme conditions.

\clearpage	
\section{Methods}

\subsection{Experiment}	

The pump-probe experiment was performed at the Linac Coherent Light Source (LCLS)~\cite{emma2010first} in Stanford, USA at the X-ray Pump Probe (XPP) beamline \cite{chollet2015x} (see also for the details of experiment~\cite{mukharamova2017probing}).
LCLS was operated in the Self Amplified Spontaneous Emission (SASE) mode.
We used LCLS in the monochromatic regime with the photon energy of a single XFEL pulse of 8~keV (1.5498~$\AA$), energy bandwidth $\Delta E / E$ of $4.4 \cdot 10^{-5}$, and pulse duration of about 50~fs at a repetition rate of 120~Hz.

The X-ray beam was focused using the Compound Refractive Lenses (CRL) on the sample down to 50~$\mu$m Full Width at Half Maximum (FWHM)
The experimental setup is shown in Fig.~\ref{Experiment} and the detailed description is given in~\cite{mukharamova2017probing}.
Series of X-ray diffraction images were recorded using the Cornell-SLAC Pixel Array Detector (CSPAD) megapixel X-ray detector~\cite{hart2012cspad}  with a pixel size of $110\times110$~$\mu$m$^2$ positioned at the distance of 10 m and covering an area approximately $17\times17$~cm$^2$.
Our experimental arrangement provided a  resolution of 0.5~$\mu$m$^{-1}$ per pixel in reciprocal space.

The Ti:sapphire IR laser was used to pump the colloidal crystals.
The pump pulses were generated at the wavelength $\lambda = 800$~nm (1.55~eV) and duration about 50~fs (FWHM).
The IR laser pulses were propagating collinear with XFEL pulses and were synchronized with the XFEL pulses with less than 0.5~ps jitter.
The size of the laser footprint on the sample was 100~$\mu$m (FWHM) and hence twice the size of the X-ray beam.

In order to obtain sufficient statistics of the measured data, for each time delay 100~diffraction patterns without IR laser and one diffraction pattern with IR laser were measured.
For two lower IR laser intensities 5 diffraction patterns were measured for each time delay.
For higher IR laser intensity measurements were repeated 9~times for 6.5~ps time delay and 10~times for 25.25~ps time delay.

\subsection{Data analysis}
Due to the varying intensity of each incoming X-ray pulse,  normalization of the diffraction patterns by the incoming beam intensity was necessary.
In order to obtain more careful characterization of FWHM of the Bragg peaks, projections on azimuthal and radial directions were performed.
These data were fitted with the one-dimensional Gaussian functions and the integrated intensity as well as broadening in radial  and azimuthal directions were determined.
	
In order to compare the dynamics of the collected data as a function of the time delay $\tau$ the following dimensionless parameters were used:
	
\begin{equation}
	\label{FWHM}
	\frac{\Delta I(\tau)}{I}=\frac{\langle I_{on}(\tau)-\overline{I_{off}(\tau}) \rangle }{\overline{I_{off}(\tau)}} \ ,
	\end{equation}
	\begin{equation}	
	\frac{\Delta w(\tau)}{w}=\frac{\langle w_{on}(\tau)-\overline{w_{off}(\tau)} \rangle }{\overline{w_{off}(\tau)}} \ .
\end{equation}
Subscript letters 'on' and 'off' define measurements with and without IR laser, respectively.
The 'off' pulses were averaged over 100~incoming pulses for each time delay.
Brackets $\langle \ldots\rangle$ correspond to averaging of the chosen Bragg peak parameter over the different positions at the sample.

\subsection{PIConGPU simulations}

To simulate plasma formation in the colloidal crystal during the first 1~ps  of the IR laser pulse propagation we used PIConGPU code~\cite{PIConGPU2013, burau2010picongpu} developed in Helmholtz-Zentrum Dresden-Rossendorf.
PIConGPU is a fully-relativistic, open-source Particle-in-Cell (PIC) code running on graphics processing units (GPUs).
The PIC algorithm solves the so-called Maxwell-Vlasov equation describing the time evolution of the distribution function of a plasma consisting of charged particles (electrons and ions) with long-range interaction.
The simulated volume of the colloidal crystal was considered according to the colloidal particle size ($d=163$ nm).
The simulation box was  $284 \times 163 \times 1150$ nm$^3$ in the $x \times y \times z$ directions  with 2.2 nm cell size in the x and z directions and 2.5 nm cell size in the y direction (see Fig.~S1 in Appendix).
On the top and bottom of the simulation box additional absorbing layers were introduced.
In the PIConGPU simulations the IR laser wavelength was set to 800~nm.
Simulations were performed with 4.25 attosecond time increment in order to resolve the plasma frequency oscillations.

\subsection{Shock wave and ablation simulations}

Shock wave and ablation simulations were performed  using 1D HELIOS code solving one-dimensional Lagrangian hydrodynamics equations.
We used two-temperature model option for the hydrodynamic simulations.
The PROPACEOS tables were used as an equation of state  for polystyrene.
The hydrodynamic simulations were coupled to the plasma PIConGPU simulations in the following way.
The 1D projection of the electron energy density profile at 1~ps obtained from the PIConGPU simulations was calculated (see Fig.~\ref{Pressure_time_z}).
The electron energy density for the initial condition was converted to electron temperature according to PROPACEOS equation of state (see Appendix for details).
The electron temperature was extended up to 6~$\mu$m according to the room temperature conditions.
The ions were assumed to have room temperature.
Calculated electron temperature and 1D projection of the density of the hexagonal-close-packed colloidal crystal structure were used as an initial condition of the hydrodynamic simulation.

Hydrodynamic simulations were performed from 1 to 1000~ps with 1~ps time increment for the 6~$\mu$m thick polystyrene colloidal crystals.
The boundaries of the plasma were allowed to expand freely.
The quiet start temperature was set to 0.044~eV which is equal to the polystyrene melting temperature.
The time increment in the HELIOS simulations was chosen according to Courant condition, and other criteria which constrain the fractional change of various physical quantities used in the simulation.
The simulation results were saved each 1~ps due to a huge amount of the output data.

\subsection*{Acknowledgements}
We acknowledge support of the project and discussions with  E. Weckert.
We acknowledge the help and discussions with T. Gurieva, Z. Jurek , A. Rode,  A. G. Shabalin,  E. A. Sulyanova, O. M. Yefanov, and L. Gelisio for the careful reading of the manuscript.
The experimental work was carried out at the Linac Coherent Light Source, a National User Facility operated by Stanford University on behalf of the U.S. Department of Energy, Office of Basic Energy Sciences.
A. Huebl and M. Garten acknowledge support from the European Cluster of Advanced Laser Light Sources (EUCALL) project which has received funding from the European Union’s Horizon 2020 research and innovation programme under grant agreement No. 654220. 
This work was supported by the Helmholtz Association’s Initiative and Networking Fund and the Russian Science Foundation, grant No. HRSF-0002.

\clearpage

\begin{table}[]
\setlength{\tabcolsep}{10pt}
\caption{ IR laser parameters and results of the plasma and shock wave simulations.
The corresponding plasma pressure was calculated from the PIConGPU simulations. Ablation depth and shock wave time and depth were calculated from the hydrodynamic HELIOS simulations. The experimental shock wave times were obtained from exponential fits of the measured data.}
\begin{tabular}{|l|c|c|c|}
\hline
Intensity, 10$^{14}$ W/cm$^2$  & 3.0 & 4.8 & 6.3 \\ \hline
Laser fluence, J/cm$^2$              & 16 & 25.5 & 33.6\\ \hline
Short times of intensity decay, ps & - & 3 $\pm$ 10 & 7.9 $\pm$ 1.1\\ \hline
Long times of intensity decay, ps & 299 $\pm$ 33 & 300 $\pm$ 28 & 275 $\pm$ 28\\ \hline
Times of radial FWHM growth, ps & 302 $\pm$ 32  & 279 $\pm$ 50 & 425 $\pm$ 90 \\ \hline
Times of azimuthal FWHM growth, ps & 345 $\pm$ 235 & 353 $\pm$ 86 & 410 $\pm$ 78 \\ \hline
Maximum electron energy density at 80~fs, $10^9$ J/m$^3$ & 49 & 98 & 149\\ \hline
Maximum electron energy density at 1~ps, $10^9$ J/m$^3$ & 12  & 20  & 25  \\ \hline
Ablation depth, nm             & 180 & 280 & 450  \\ \hline
Shock wave depth, $\mu$m        & 2.36 & 4.15 & 5.00   \\ \hline
Maximum mass velocity, km/s        & 2.3 & 2.7 & 2.8   \\ \hline
Simulated shock wave stop times, ps  & 437 & 756 & 931 \\ \hline

\end{tabular}
\label{Table}
\end{table}

\clearpage

\clearpage

\renewcommand{\thefigure}{S\arabic{figure}}
\renewcommand{\thetable}{S\arabic{table}}
\setcounter{figure}{0}
\setcounter{section}{0}
\setcounter{table}{0}
{\bf{\Large{Appendix}}}

\section{Infrared laser calibration}

\begin{figure}[h]
	\includegraphics[width=0.9\linewidth]{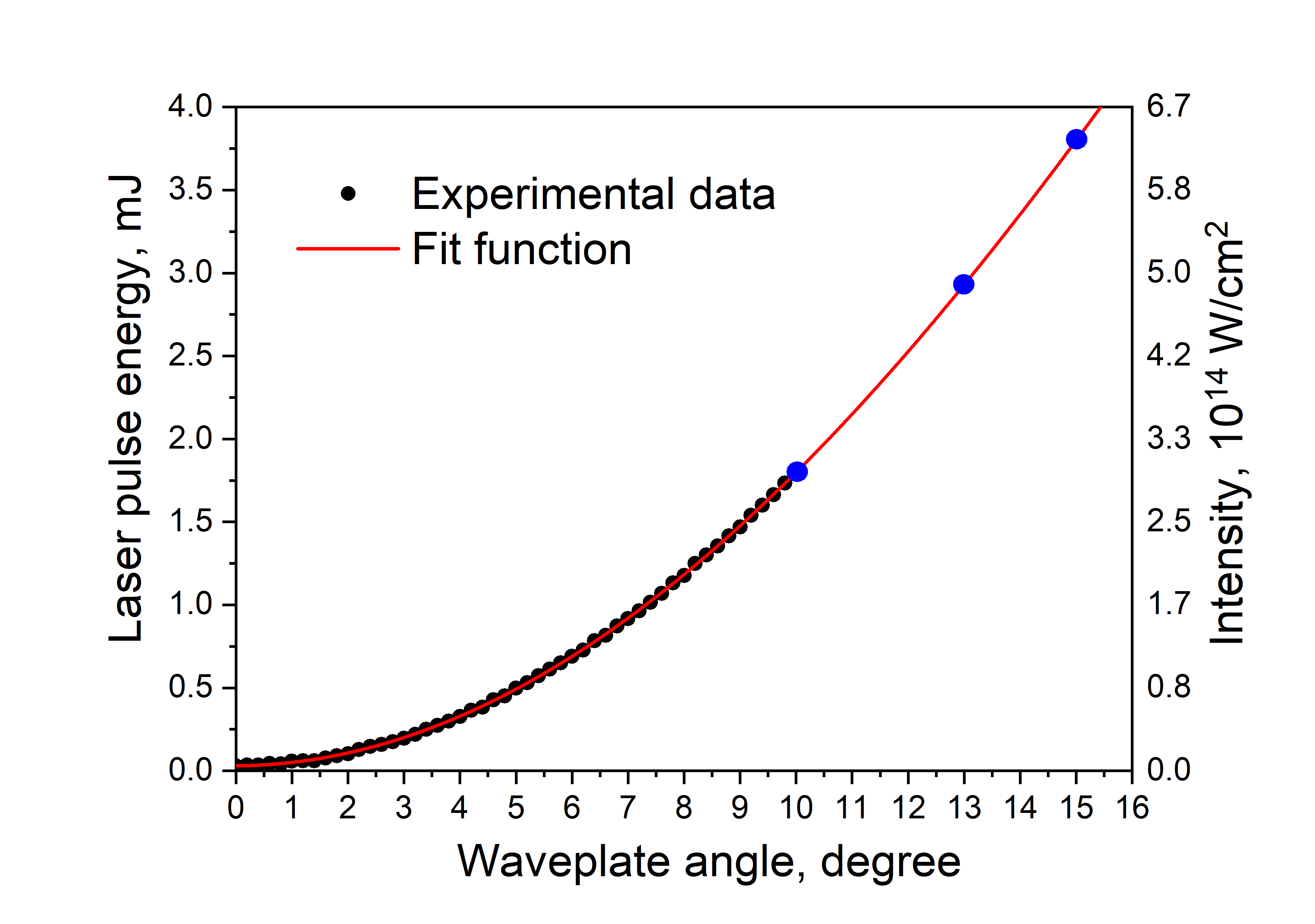}
	\caption{
		The infrared (IR) laser energy calibration curve. The measured data is shown by the black dots and the sine fit is shown by the solid red line. The  IR laser energy used in the experiments for three samples is marked with blue circles.	}
	\label{Calibration}
\end{figure}

In our pump-probe experiment the Ti:sapphire IR laser was used to pump the colloidal crystal film.
The IR laser energy was controlled by the rotation of the optical axis of a waveplate, and was calibrated by power sensor at the position of the sample.
The  calibration curve showing the dependence of the laser pulse energy from the waveplate angle is presented in Fig.~\ref{Calibration}.
The corresponding IR laser intensity is shown on the right vertical axis.
The IR laser intensity was calculated from the IR laser energy assuming Gaussian shape of the pulse with 50~fs FWHM in the temporal domain and 100~$\mu$m FWHM in the spatial domain.
Zero degrees of waveplate angle corresponds to the minimum and 15~degrees correspond to the maximum calibrated energy and intensity of the IR laser.
The calibration curve was fitted with a sine function shown by the solid red line in Fig.~\ref{Calibration}.
The three IR laser intensities ($I_1=3.0 \cdot 10^{14}$ W/cm$^2$, $I_2=4.8 \cdot 10^{14}$ W/cm$^2$ and $I_3=6.3 \cdot 10^{14}$ W/cm$^2$) used in the current experiment are marked by the blue circles.
At energies lower than 1 mJ, no ultrafast melting  was observed and different dynamics of the colloidal crystal was investigated in a separate work~\cite{mukharamova2017probing}.

\section{Plasma formation simulations}

\subsection{Simulation cell}

\begin{figure}
	\includegraphics[width=0.95\linewidth]{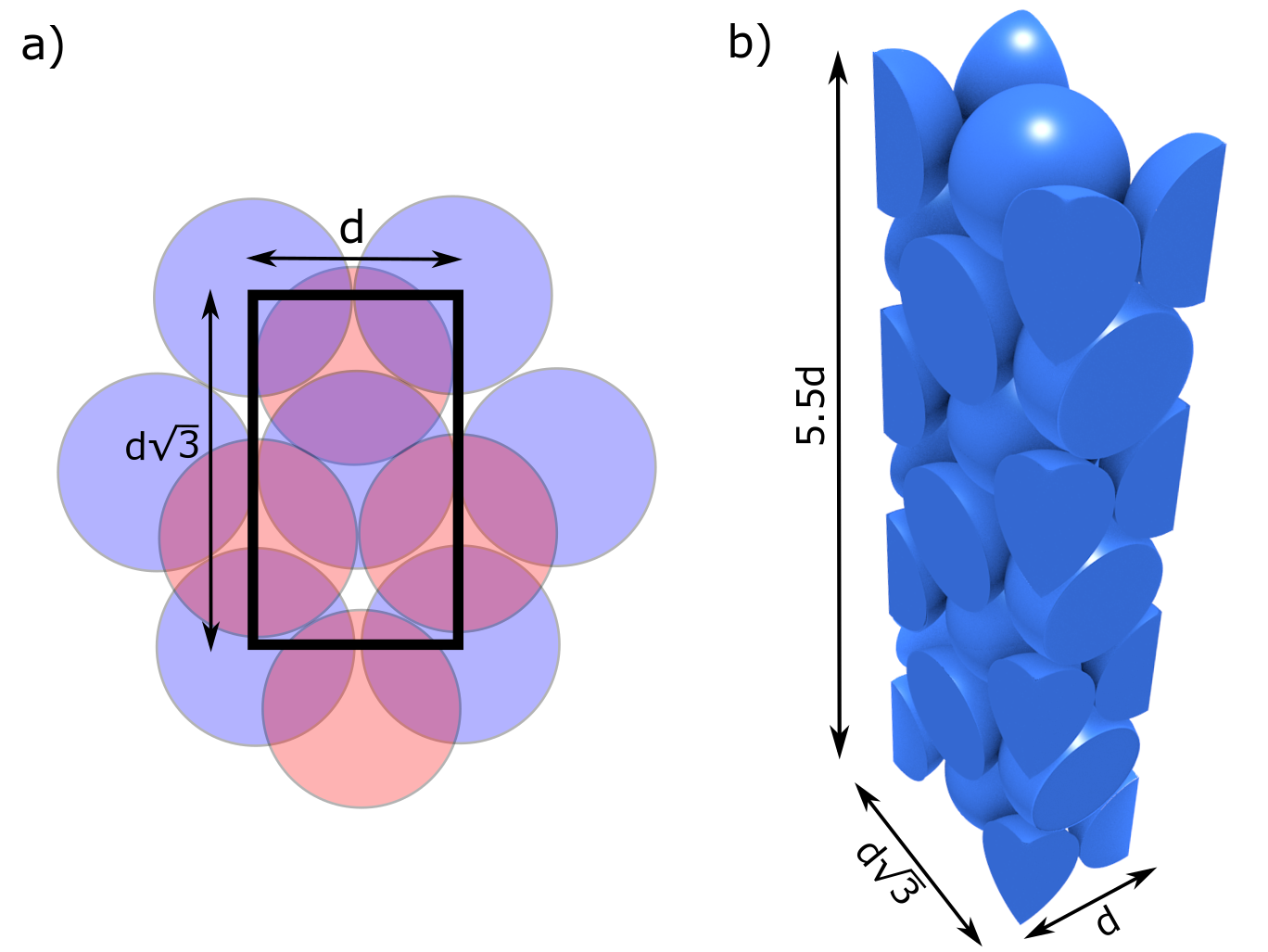}
	\caption{
		Simulation cell used in PIConGPU simulations. a) The top view on the simulation box. Different layers of the hexagonal-close-packed colloidal particles are shown in blue and red color. b) 3D view on the simulation box.  	}
	\label{SimCell}
\end{figure}

To simulate plasma formation in the colloidal crystal during the first 1~ps  of the IR laser pulse propagation we used PIConGPU code version 0.4.0--dev developed at Helmholtz-Zentrum Dresden-Rossendorf~\cite{burau2010picongpu,PIConGPU2013}.
In the PIConGPU simulation a rectangular shape of the simulation box is considered.
The simulated volume of the colloidal crystal was chosen according to the colloidal particle size (d$ =163$ nm).
The simulation box is shown in Fig.~\ref{SimCell} and was  $ 284 \times 163 \times 1280 $ nm$^3$ in $x \times y \times z$ directions ($d  \times d \sqrt3$ in $x \times y$ direction).
The top view on the simulation box is shown in  Fig.~\ref{SimCell}(a) and two layers of the hexagonal-close-packed colloidal crystal are shown by red and blue color.
From Fig.~\ref{SimCell}(a) it is clear that such a simulation box is periodic in x and y direction.
Therefore, such a size of the simulation box was chosen to  apply  periodic boundary conditions on the sides (x=0 nm, x = 284 nm and y=0 nm, y=163 nm planes) of the simulation box.
On the top and bottom of the simulation box additional absorbing layers were introduced.
To optimize the simulation process the simulation box consisted of $128 \times 64 \times 512$ cells in $x \times y \times z$ directions.
From that condition, the size of one cell was chosen to be 2.2 nm in x  and 2.5 nm in y and z directions.
On the top of the simulation box 128 cells or 320 nm were  not filled with any colloidal particles.
This empty space was introduced in the simulation box to initialize the incoming IR laser.

In order to satisfy Courant Friedrichs Lewy  condition~\cite{courant1928partial} the simulations were performed with 4.25 as time increment.
Such a time increment allows to resolve the plasma frequency oscillations in the $2.2 \times 2.5 \times 2.5 \times$ nm cell size  in $x \times y \times z$ directions.
The output of the simulation was saved each 5~fs during the first 100~fs and each 20~fs up to 1~ps due to a huge  amount of the output data.
We used a standard Yee solver scheme~\cite{yee1966numerical} and the HDF5 openPMD output~\cite{huebl_axel_2015_33624}  implemented in the PIConGPU code.

\subsection{ADK ionisation of the colloidal crystal}

\begin{figure}
	\includegraphics[width=\linewidth]{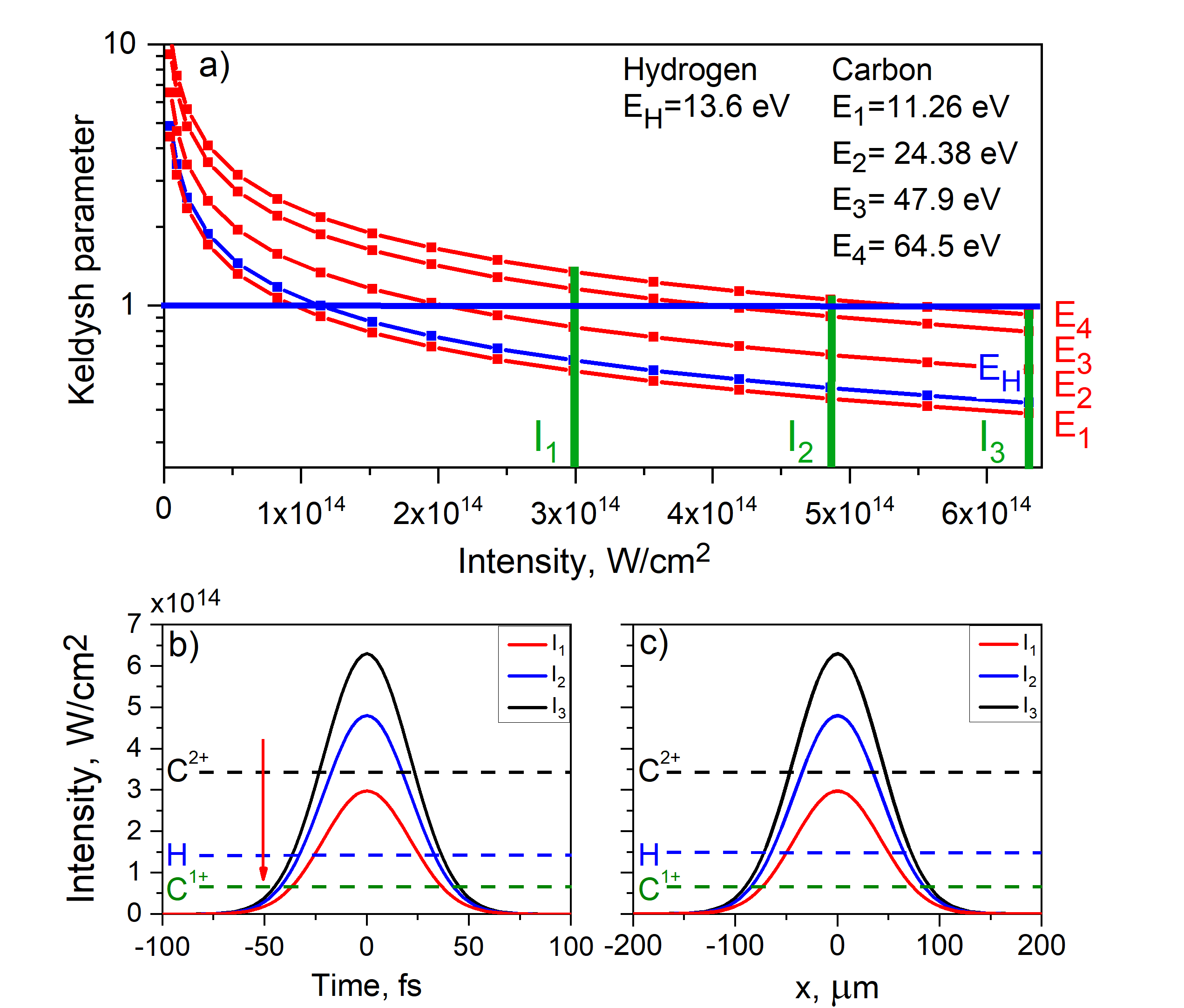}
	\caption{a) Keldysh parameter for 4 carbon ionisation energies ($E_1-E_4$) and hydrogen ionisation energy $E_H$. Measured intensities $I_1=3.0~\cdot 10^{14}$~W/cm$^2$, $I_2=4.8~\cdot 10^{14}$~W/cm$^2$ and $I_3=6.3~\cdot 10^{14}$~W/cm$^2$ are shown by green vertical lines. For C$^1+$, C$^2+$ and H$^+$ Keldysh parameter is lower than one for all three intensities. Temporal b) and  spatial c)  Gaussian profile of the IR laser. ionisation thresholds for C and H are indicated by horizontal dashed lines. Beginning of the simulation is indicated as vertical red arrow. }			
	\label{KeldyshParam}
\end{figure}

Field ionisation process can be described according to Keldysh theory~\cite{keldysh1965ionization}.
For  low fields and high frequencies the Keldysh parameter $\gamma>1 $ and multi-photon ionisation happens while for strong fields and low frequencies $\gamma<1$ and tunneling ionisation prevails.
Dependence of the Keldysh parameter on the IR laser intensity is shown in Fig.~\ref{KeldyshParam}(a).
It is clearly seen that for three IR laser intensities used in our experiment Keldysh parameter is lower than one so quasi-static ionisation regime occurred as is also shown in Refs. \cite{reiss2014tunnelling, reiss2008limits}.

Depending on the electric field $E$ generated by the laser in the quasi-static regime, ionisation can be described as tunneling or above-barrier ionisation (ABI).
If laser field energy is higher than the threshold $E_{ABI}=E_{i}/4Z$, where $(Z-1)$ is ion charge, the ionisation is above barrier and if it is lower the tunneling ionisation is prevailing~\cite{krainov2002cluster}.
ionisation thresholds for ABI are shown at the temporal and spatial profiles of the IR laser intensity in Fig.~\ref{KeldyshParam}(b,c) by horizontal dashed lines.
Beginning of the simulation time is indicated in Fig.~\ref{KeldyshParam}(b) by the vertical red arrow 50 fs before the IR laser pulse reaches it's maximum.
It is well seen that at the beginning of the simulation the IR laser intensity is below ionisation thresholds for C and H.

From Fig.~\ref{KeldyshParam}(b,c) it is well seen that for lower laser intensity of $3 \cdot 10^{14}$ W/cm$^2$ only C$^{1+}$ and H$^{1+}$ are expected to be fully ionized.
For two higher laser intensities we also have ionisation of C$^{2+}$.
It is interesting to look also at the spatial distribution of the IR laser pulse (see Fig.~\ref{KeldyshParam}(c)).
In the area of about 150~$\mu$m the sample is ionized to C$^{1+}$ and H$^{1+}$ for lower laser intensity, and to C$^{2+}$ for two higher IR laser intensities.
Therefore in the 50~$\mu$m focused x-ray beam the plasma was considered to be ionized.

The ionisation rate $\Gamma$ implemented in PIConGPU code was calculated according to Ammosov-Delone-Krainov (ADK) model ~\cite{delone1998tunneling}  in the case of a linearly polarized field (an s-state is taken for simplicity) as
\begin{equation}
\Gamma_{ADK}=\sqrt{\frac{3n^{*3}E}{\pi Z^3}} \frac{E}{8 \pi Z} \bigg( \frac{4 e Z^3}{E n^{*4}} \bigg)^{2n*} \exp \bigg(-\frac{2Z^3}{3n^{*3}E}\bigg),
\label{Eq:ADK}
\end{equation}
where $n^*=Z/\sqrt{2E_{i}}$ is the effective principal quantum number.
The ionisation equations are only in this form if you use the atomic unit system.
The ionisation probability is calculated from the ionisation rate as $P=1-e^{-\Gamma_{ADK}\Delta t}$.

The version of the ADK model that was used in the PIConGPU code is simplified due to the fact that our particles are carbon and hydrogen and do not have much inner structure. 	
This model was applied for both the tunneling regime $E<E_{ABI}$ and above-barrier regime $E \approx E_{ABI}$.
For strong fields $E>E_{ABI}$ where the potential barrier binding an electron is completely suppressed the so-called barrier-suppression ionisation (BSI) regime is reached.
Therefore,  the ADK model was combined  with a check for the BSI threshold.
Also the ADK model used in this work  is based on the assumption that the material investigated consists of independent atoms and it did not consider the molecular-orbital structure of covalently bonded materials.

\begin{figure}
	\includegraphics[width=\linewidth]{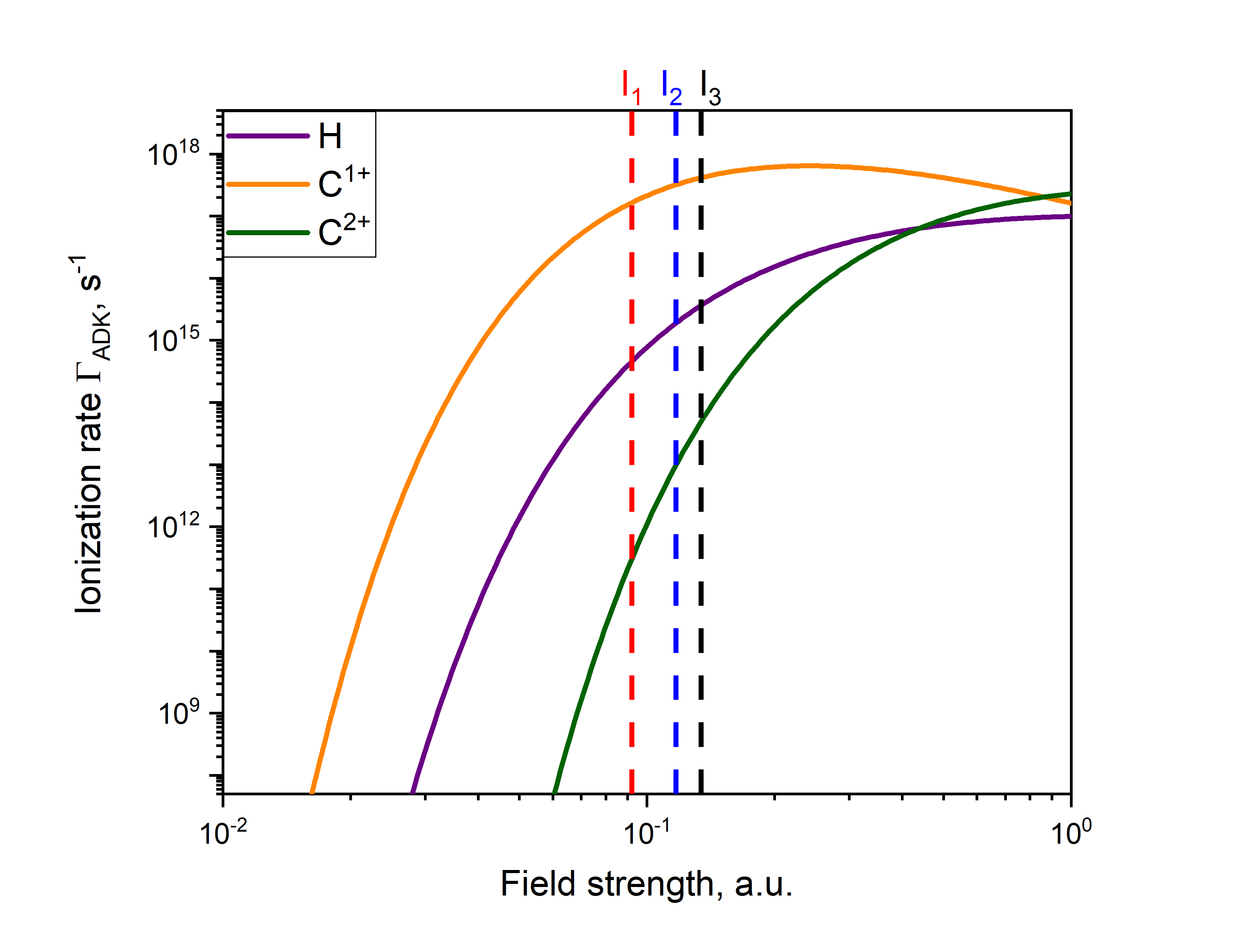}
	\caption{ADK ionisation rates for three different IR laser intensities. The ionisation rates are shown for H, C$^{1+}$ and C$^{2+}$ by solid lines. IR laser intensities I$_1=3.0 \cdot 10^{14}$ W/cm$^2$,  I$_2=4.8 \cdot 10^{14}$ W/cm$^2$, I$_3=6.3 \cdot 10^{14}$ W/cm$^2$ are marked by the vertical dashed lines. }		\label{ADK}
\end{figure}

In Fig.~\ref{ADK}  ionisation rates calculated according to ADK ionisation model for H, C$^{1+}$ and C$^{2+}$ are shown.
The intensities of the IR laser used in our experiment are marked by the vertical dashed lines.
As can be seen from Fig.~\ref{ADK} the ionisation rate is growing with the field strength up to $10^{17}$ s$^{-1}$ for the C$^{1+}$.
From this figure one can conclude that  each carbon and hydrogen atoms are ionized up to C$^{1+}$ and H$^{1+}$ for all three IR laser intensities and for two higher IR laser intensities significant amount of C$^{2+}$ is created.

\subsection{Thomas-Fermi ionisation of the colloidal crystal}

The second ionisation mechanism is collisional ionisation.
Accelerated electrons collide inelastically with the atomic ions inside the  colloidal particle which causes ionisation of the atom.
Due to the collisional ionisation mechanism, the atoms in the inner part of the colloidal particle are ionized to C$^{4+}$ for the highest IR laser intensity.
The ionisation rate of this process was calculated according to the Thomas-Fermi ionisation model~\cite{more1985pressure}.
Thomas-Fermi ionisation model uses the self-consistent method, where atom is represented as a point
nucleus embedded in a spherical cavity in a continuous background positive charge.
The cavity radius $R_0$, is determined by the plasma density ($\rho= 3M_p/4 \pi R_0^2$, where $M_p$ is the atomic mass).
The ionisation state was calculated using an approximate fit to the definition of the ionisation state
\begin{equation}
Z^*(\rho, T)=4/3 \pi R_0^3 n(R_0),
\label{Eq:ADK}
\end{equation}
where the ionisation state $Z^*(\rho, T)$ is defined from the  boundary density $n(R_0)$.
The parameters of the fit which were used in PIConGPU can be found in Table~4 in Ref.~\cite{more1985pressure}.
The ion proton number, ion species mass density, and electron temperature are used as an input for the Thomas-Fermi ionisation model in the PIConGPU simulation.

\begin{figure}
	\includegraphics[width=\linewidth]{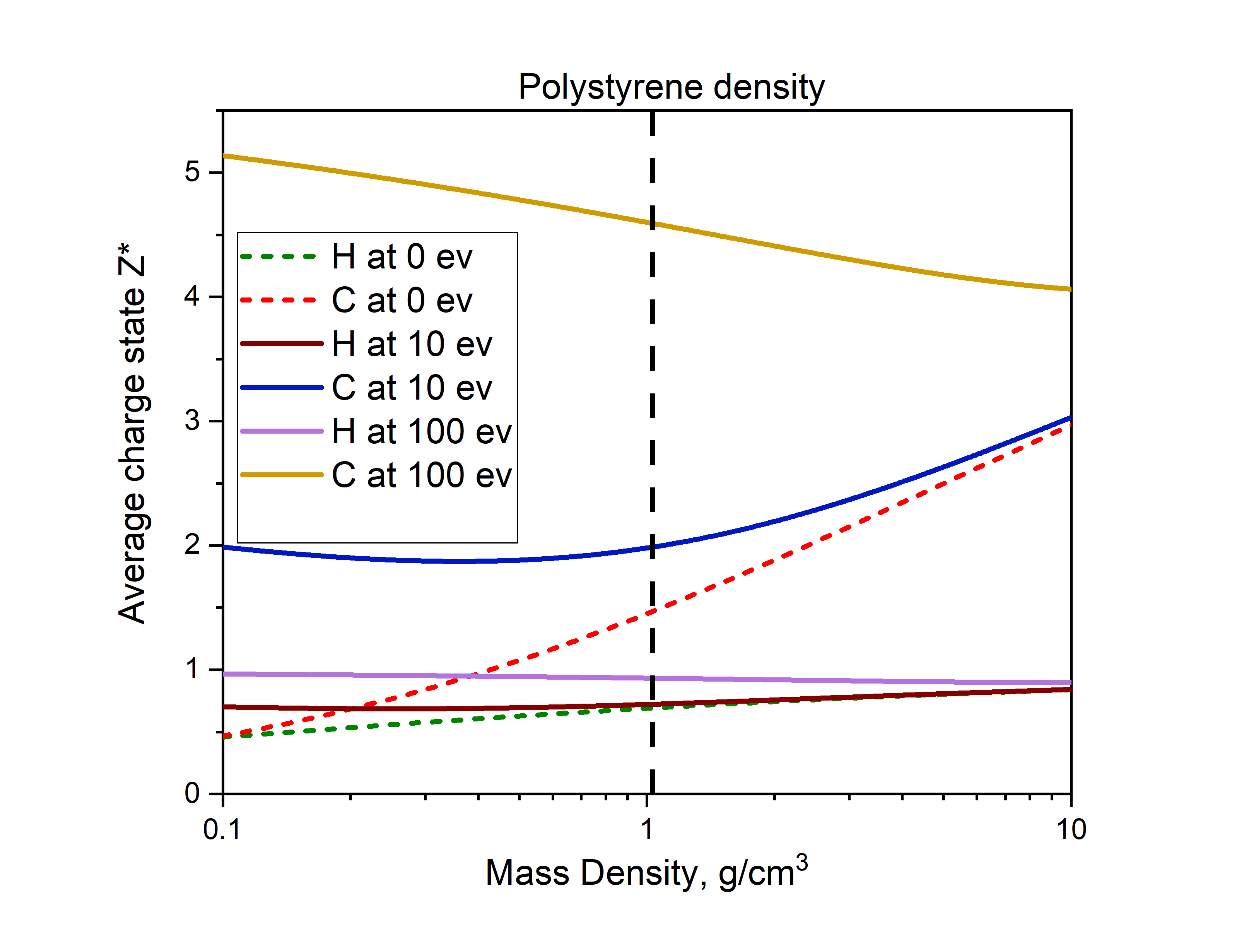}
	\caption{Thomas-Fermi ionisation for H and C for 0~eV, 10~eV and 100~eV electron temperatures.
		For 10~eV and 100~eV charge state predictions are marked by solid lines and for 0~eV the charge state predictions are marked by dashed lines because they show unphysical behavior and were excluded from the simulations.}			
	\label{TF}
\end{figure}

The charge state estimates for carbon and hydrogen obtained from this model are shown in Fig.~\ref{TF}.
As can be seen from this figure the Thomas-Fermi model displays unphysical behavior in several cases,  thus the cutoff values were introduced, to exclude some particles from the calculation.
For carbon and hydrogen  it predicts non-zero charge states at zero temperature, therefore the lower electron-temperature cutoff value should be defined, and in our model it was 1~eV.
For low ion densities Thomas-Fermi model predicts an increasing charge state for decreasing ion densities (see Fig.~\ref{TF}).
This occurs already for electron temperatures of 10~eV and the effect increases as the temperature increases.
Low ion-density cutoff value was $1.74 \cdot 10^{21}$~ions/cm$^3$ in our case.
Also, super-thermal electron cutoff value was introduced to exclude electrons with kinetic energy above 50~keV.
That is motivated by a lower interaction cross-section of particles with high relative velocities.


\subsection{Ionisation simulations}

\begin{figure}
	\includegraphics[width=\linewidth]{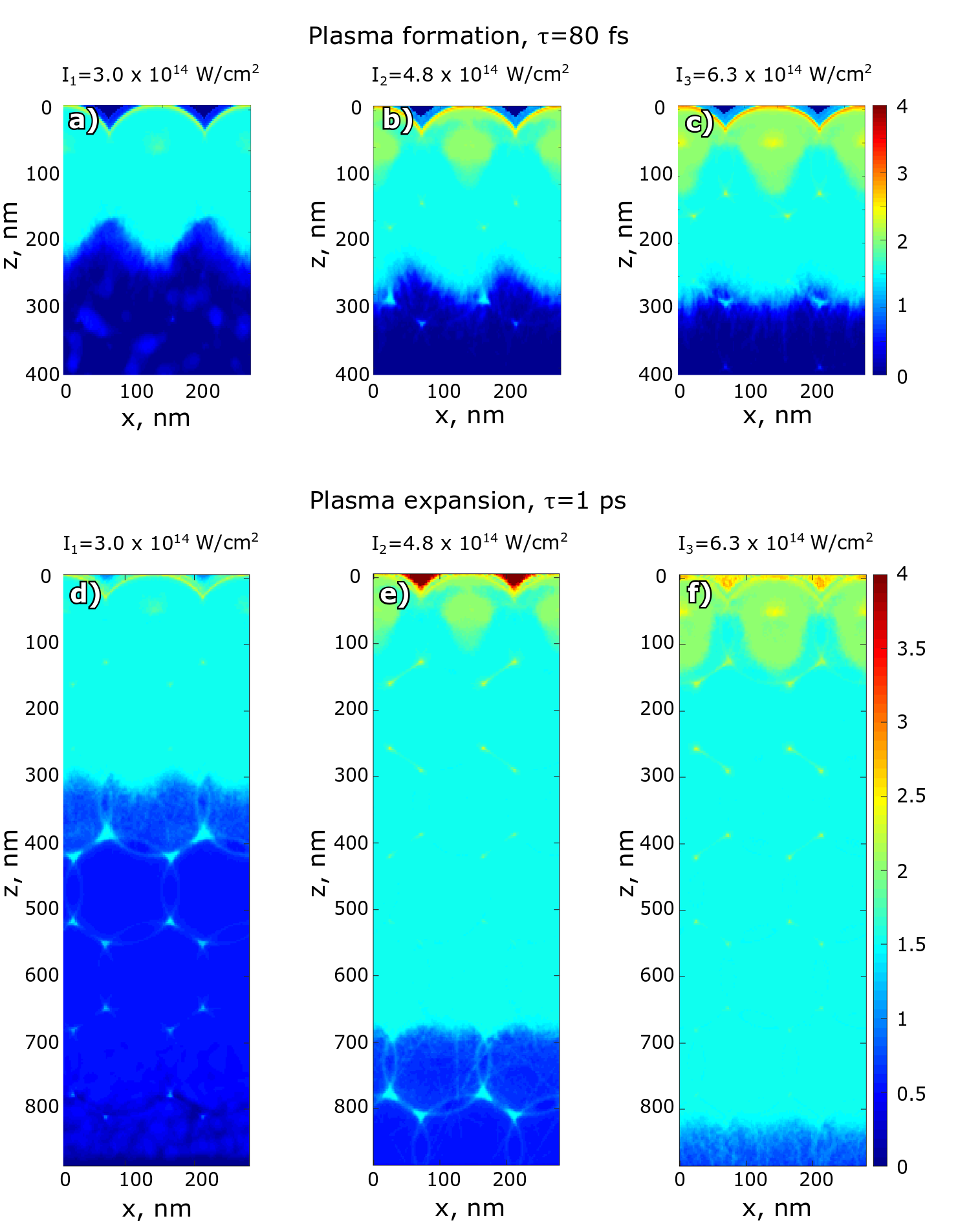}
	\caption{ The average charge state distribution in the colloidal crystal at 80 fs (a-c) and at 1~ps (d-f) after the beginning of the IR laser pulse for three different IR laser intensities. Here we show the projection of the average charge state along the y-direction. }			
	\label{Charge}			
\end{figure}

PIConGPU simulations were performed for all three IR laser intensities measured in our pump-probe experiment.
PIConGPU simulations intrinsically included field and collisional ionisation discussed in the previous section and thus the averaged charge state of the colloidal crystal was obtained.
The average charge state projections along the y-axis after 80~fs and after 1~ps of the IR laser pulse propagation are shown in Fig.~\ref{Charge}.
The IR laser pulse is coming from the top along the z-direction, and it starts the ionisation of the colloidal crystal sample.

As one can see from Fig.~\ref{Charge}(a-c), after 80~fs only the first layer of the colloidal particles is ionized by the IR laser pulse.
The highly ionized skin layer of the thickness about 10~nm on top of the first layer of colloidal crystals is also well visible.
The average charge state in the skin layer can reach up to 2.9 and is summarized in Table~\ref{Table}.
Also, high charge state is reached in the center of the colloidal particles for all three IR laser intensities, and their values are summarized in Table~\ref{Table}.

At 600~fs after the beginning of the simulation accelerated electrons collisionally ionized the inner part of the colloidal crystal and until 1~ps the ionisation state of the colloidal crystal remained practically constant (see Fig.~\ref{Charge}(d-f) and Supplementary movie).
From Fig.~\ref{Charge}(d-f) it can be observed that the ionisation depth varies a lot with the IR laser intensity.
The ionisation depth is different for three IR laser intensities and is summarized in Table~\ref{Table}.
Even at 1~ps after the start of the PIConGPU simulation the highest ionisation state remains at the center of the top layer of the colloidal particles.

\begin{table}[b]
	\setlength{\tabcolsep}{10pt}
	\caption{
		IR laser parameters and results of plasma simulations used in our experiment. The corresponding average ionisation state and ionisation depth were calculated from the PIConGPU simulations.}
	\begin{tabular}{|l|c|c|c|}
		\hline
		Intensity, 10$^{14}$ W/cm$^2$  & 3.0 & 4.8 & 6.3 \\ \hline
		Average charge state in the skin at 80~fs  & 2.1 & 2.7 & 2.9 \\ \hline
		Average charge state in the center of the first colloidal particle at 80~fs  & 1.8 & 2.1 & 2.4 \\ \hline
		Average charge state in the skin at 1~ps  & 2.1 & 2.7 & 2.7 \\ \hline
		Average charge state in the center of the first colloidal particle at 1~ps  & 1.7 & 2.1 & 2.4 \\ \hline
		Ionisation depth, nm              & 300 & 700 & 820\\ \hline
		\hline
	\end{tabular}
	\label{Table}
\end{table}

\section{Hydrodynamic simulations}

\begin{figure}[h]
	\includegraphics[width=\linewidth]{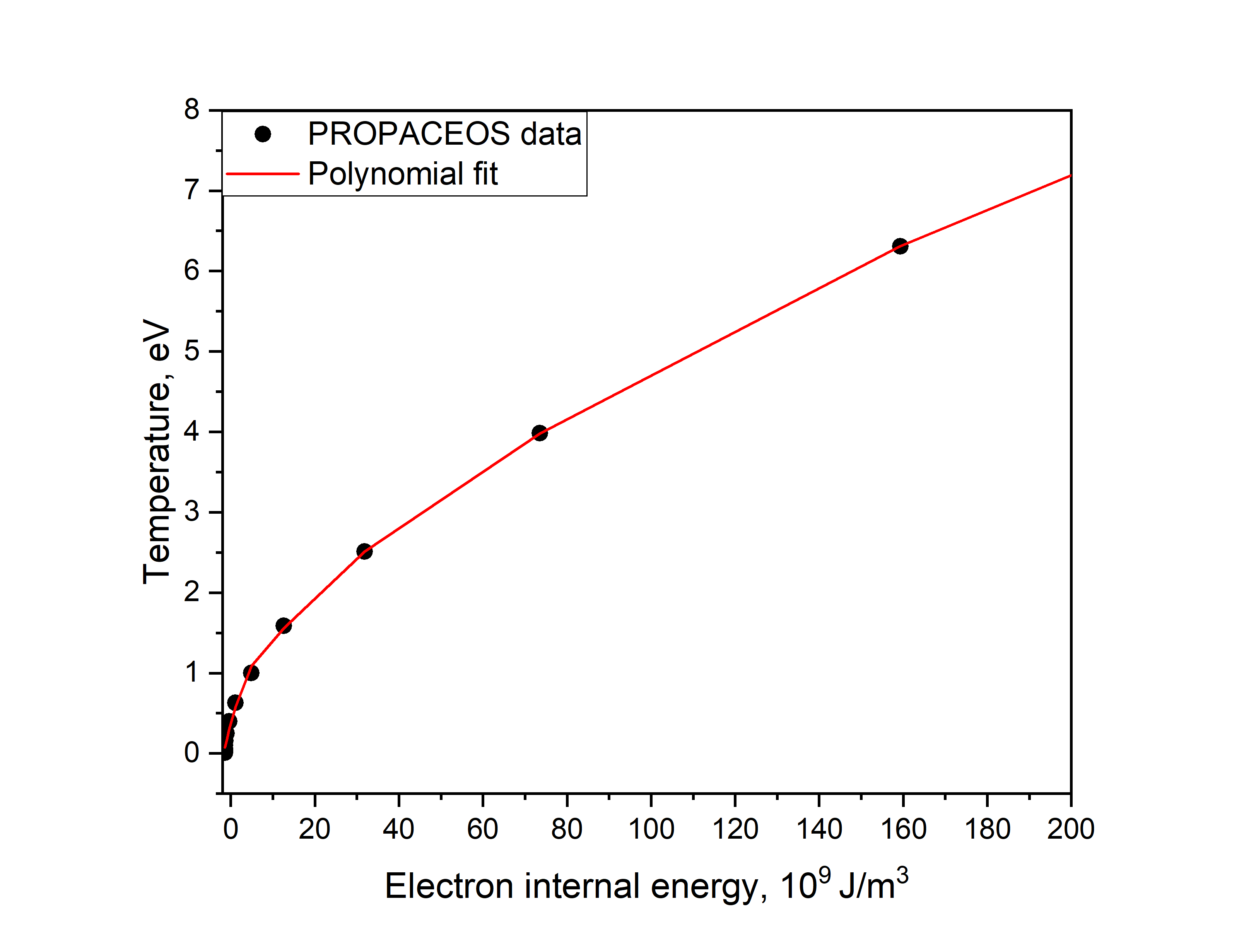}
	\caption{Polystyrene equation of state obtained from PROPACEOS data tables (black dots) and polynomial fit (solid red line).}			
	\label{PROPACEOS_EOS}			
\end{figure}

\begin{figure}[h]
	\includegraphics[width=\linewidth]{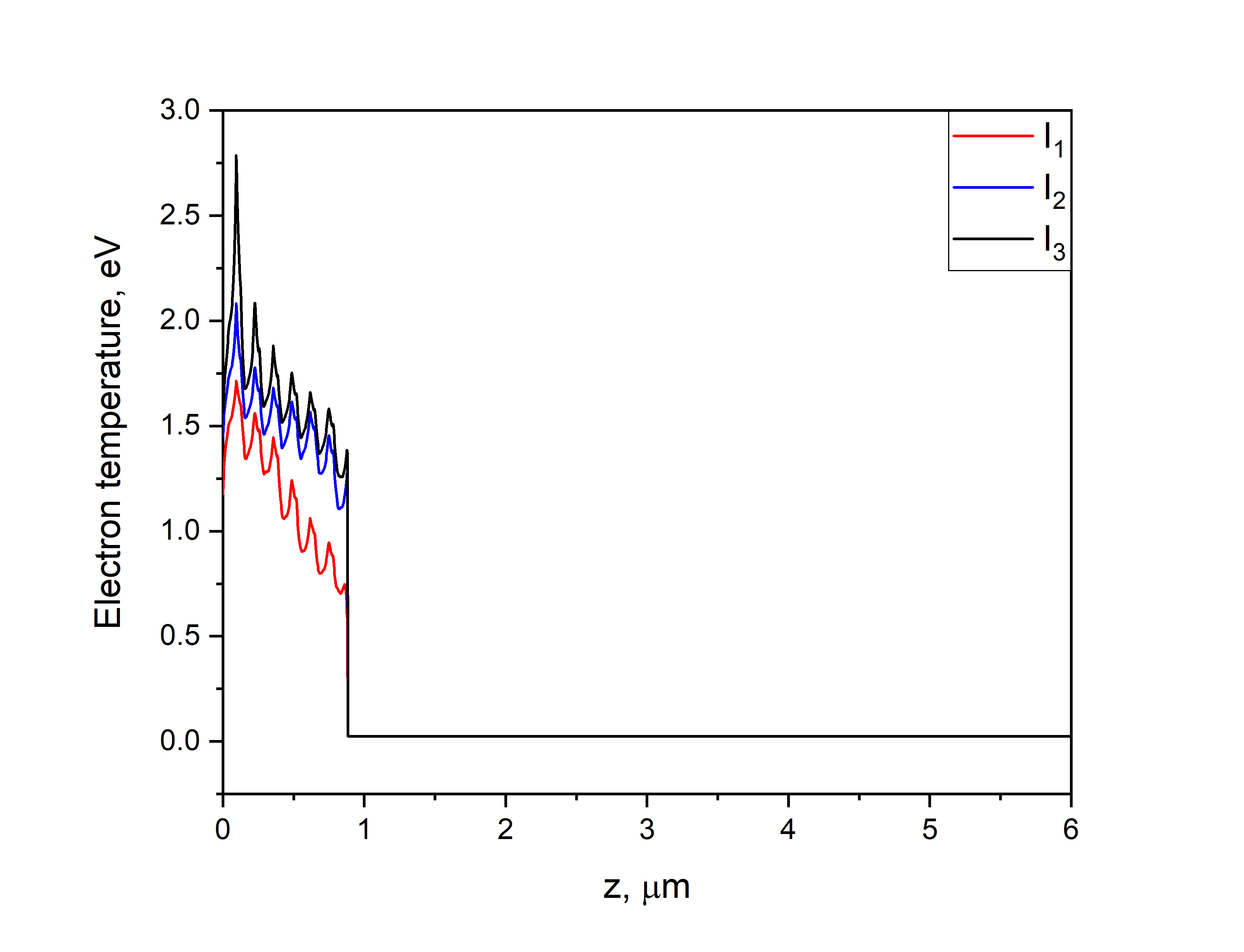}
	\caption{Electron temperature distribution in the colloidal crystal for three measured intensities $I_1=3.0 \cdot 10^{14}$ W/cm$^2$, $I_2=4.8 \cdot 10^{14}$ W/cm$^2$ and $I_3=6.3 \cdot 10^{14}$ W/cm$^2$ at 1 ps after the IR laser pulse.}			
	\label{Temp_input}			
\end{figure}

\subsection{PIConGPU simulations coupled to HELIOS simulations}

The HELIOS hydrodynamic simulations were performed to model structural changes in the colloidal crystal due to the shock wave propagation.
The hydrodynamic simulations were coupled to the plasma PIConGPU simulations in the following way.
The 1D projection of the electron energy density profile at 1~ps was obtained from the PIConGPU simulations.
The electron energy density distribution was further converted to the electron temperature using PROPACEOS (PRism OPACity and Equation Of State code) data tables.
The horizontal axis is the electron internal energy and the vertical axis is the electron temperature.
The PROPACEOS data for polystyrene with typical polystyrene density of 1.05~g/cm$^3$ is shown by the black dots.
This data was fitted with 7th order polynomial function and  the polynomial fit is shown by the red line in Fig.~\ref{PROPACEOS_EOS}.
Further this polynomial function was used to convert the electron energy density to electron temperature.

The temperature distribution used as an input for hydrodynamic simulations is shown in Fig.~\ref{Temp_input} for three IR laser intensities.
The oscillations of the electron temperature due to the periodic colloidal crystal structure are  clearly visible.
The PIConGPU simulations were performed for the first 882.5~nm of the colloidal crystal, below this depth up to 6~$\mu$m the temperature distribution was set to a room temperature value  of 0.025~eV (see Fig.~\ref{Temp_input}).
By that we got a drop of the temperature distribution at 882.5~nm depth, which was not smoothed for deeper parts of the colloidal crystal due to the following reasons.
The ionisation rate in the PIConGPU code does not take into account the recombination process, due to that the temperature distribution on the top of the sample should be lower in reality.
Our approach was later confirmed by the hydrodynamic simulations performed only by HELIOS code (see section IIIc in Appendix). 

These PIConGPU and HELIOS combined set of simulations is further referred to as a simulation Set~1.

\subsection{Results of the simulation Set~1}

\begin{figure}
	\includegraphics[width=0.85\linewidth]{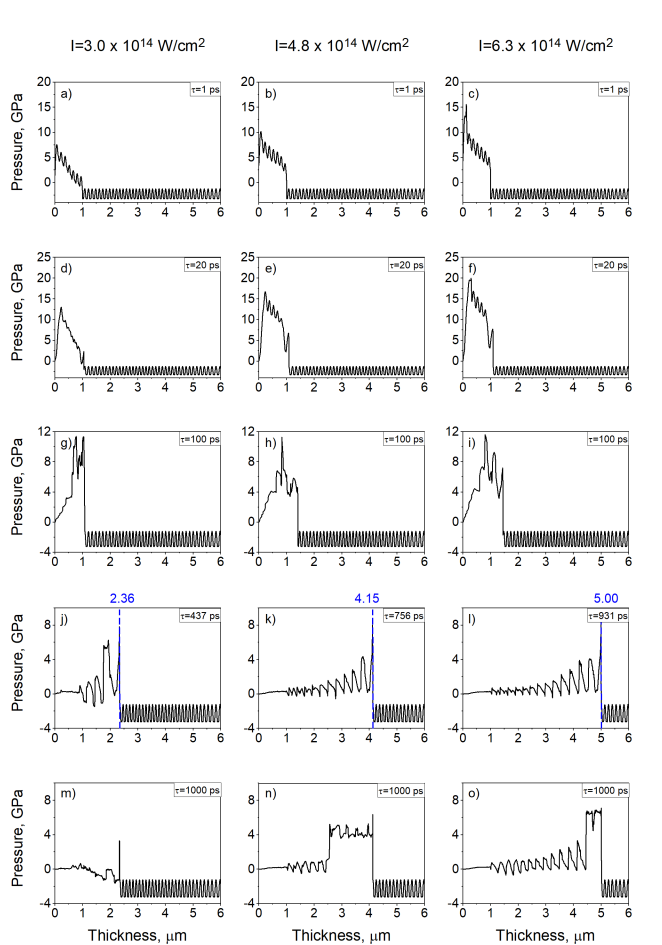}
	\caption{ Pressure in the colloidal crystal from the hydrodynamic simulation. In this case the combination of PIConGPU and HELIOS code was used.  Results are shown at 1~ps (a-c), 20 ps (d-f), 100~ps (g-i) after the stop of the shock wave (j-l) and at 1000~ps, the end of the simulation (m-o). The blue dashed line shows the shock wave stop depth (j-l). }			
	\label{Pressure_1D}			
\end{figure}

\begin{figure}
	\includegraphics[width=0.85\linewidth]{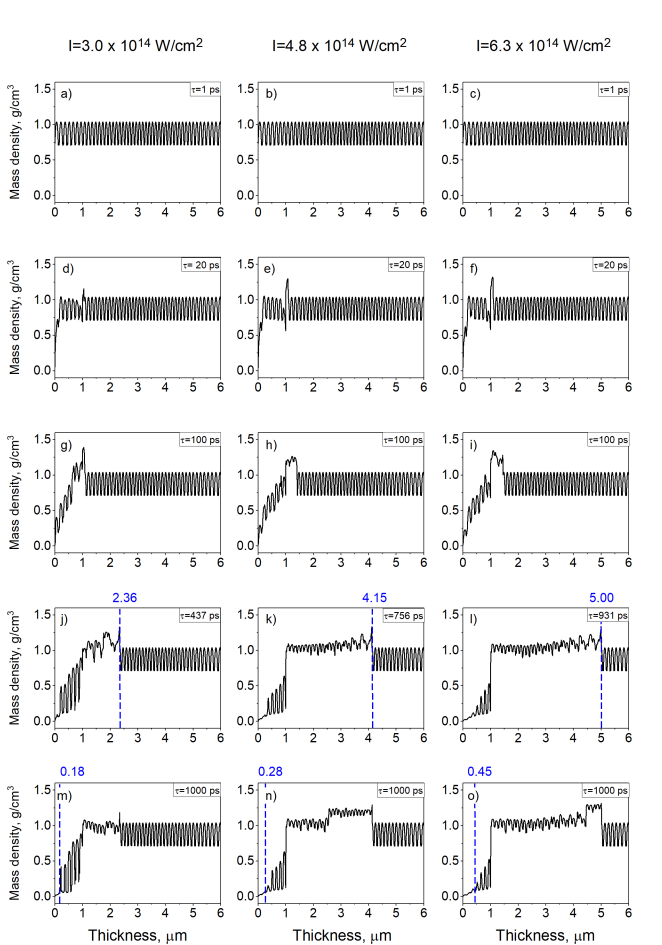}
	\caption{ Mass density of the periodic colloidal crystal from the hydrodynamic simulation. In this case the combination of PIConGPU and HELIOS code was used. Results are shown at 1~ps (a-c), 20 ps (d-f), 100~ps (g-i) after the stop of the shock wave (j-l) and at 1000~ps, the end of the simulation (m-o). The blue dashed line shows the shock wave stop depth (j-l) and the ablation threshold (d-f).}			
	\label{Mass_1D}			
\end{figure}

The 1D pressure  and mass density distribution obtained from the hydrodynamic simulations are shown in Fig.~\ref{Pressure_1D},~\ref{Mass_1D}, respectively, for all three IR laser intensities.
The initial density distribution used as an input for the hydrodynamic simulations is shown in Fig.~\ref{Mass_1D}(a-c).
These density profiles were obtained as a projection on the hexagonal-close-packed colloidal crystal along x- and y-direction.
The periodicity of the mass density due to colloidal crystal structure is well resolved with the chosen step size of 2.5~nm.
The initial pressure distributions are shown in Fig.~\ref{Pressure_1D}(a-c) as snapshots at 1~ps after the beginning of the IR laser pulse.
As one can see  at 1~ps the pressure is decaying along the z-direction, and the maximum pressure is in the center of the top layer of the colloidal particles.
In the deeper part of the sample the pressure distribution is periodic due to the periodicity of the colloidal crystal.
The negative pressures in the simulation are zeroth-order approximation to material strength used in the PROPACEOS and other, for example SESAME EOS tables.

During the first picoseconds of the shock wave propagation the mass density of the approximately 1~$\mu$m region on the top of the sample is affected.
The snapshot of the mass density at 20~ps is shown in Fig.~\ref{Mass_1D}(d-f) where the ablation of the material is clearly visible.
The modulations of the mass density become less pronounced and the steep gradient of the mass density going down to zero value is well visible.
That is a clear sign of the ablation of the top layer of the colloidal crystal.
At 20~ps after the beginning of the HELIOS simulation the high pressure from the top of the colloidal crystal is propagating inside the sample (see Fig.~\ref{Pressure_1D}(d-f)).
At the top 100~nm of the sample pressure is equal to zero due to ablation of the first layer of the colloidal crystal.
At about 100~ps the highest pressure is reaching the shock wavefront and the shock wave propagates inside the sample with increased speed.

The shock wave propagating inside the colloidal crystal compresses the surrounding material and from Fig.~\ref{Mass_1D} it is clearly visible that density at the shock wavefront is higher than density after the shock wavefront.
When the energy of the shock wave is not sufficient to compress the sample the shock wave stops.
The shock wave stops at different times for three IR laser intensities.
The pressure distribution at the moment when the shock wave stops is shown in Fig.~\ref{Pressure_1D}(j-l).
The shock wave depth is marked by the blue dashed line.
The shock wave stops at different time and depth for all three IR laser intensities (see Table~1 in the main text).
The mass density deeper the shock wavefront remains unperturbed while  before the shock wavefront is undergoing some small changes even after the shock wave stops (see Fig.~\ref{Mass_1D}(j-o)).

The last time point of the hydrodynamic simulation was at 1000~ps and the mass density and pressure snapshots are shown in Fig.~\ref{Mass_1D}(m-o) and Fig.~\ref{Pressure_1D}(m-o).
The shock wave did not propagate any deeper inside the colloidal crystal but the mass density on the shock wavefront is smaller than in Fig.~\ref{Mass_1D}(j-l) and the pressure front changed its shape significantly.
The ablation of the material has also practically finished at that time.
From Fig.~\ref{Mass_1D}(m-o) it is well seen that the amount of ablated material is higher for higher IR laser intensity, and the values are summarized in the Table~1 in the main text.
The ablation threshold is marked by the blue dashed line in Fig.~\ref{Mass_1D}(d-f).

\begin{figure}[h]
	\includegraphics[width=1.0\linewidth]{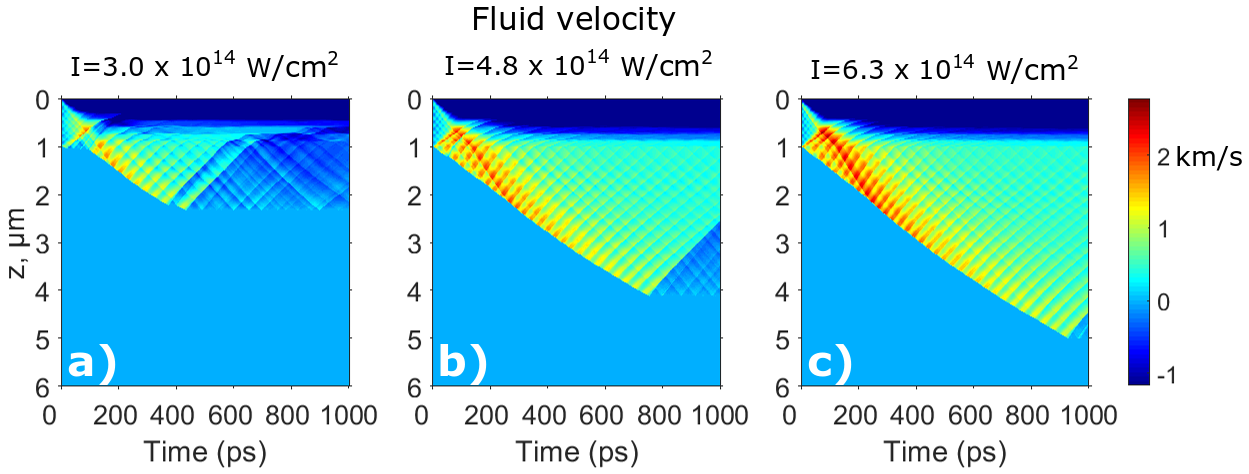}
	\caption{ Hydrodynamic simulations of the shock wave propagation. Color plots show simulation results for the fluid velocity  (a-c) for three different IR laser intensities: (a)~$I_1$=$3.0 \cdot 10^{14}$ W/cm$^2$,  (b)~$I_2$=$4.8 \cdot 10^{14}$ W/cm$^2$, (c)~$I_3$=$6.3 \cdot 10^{14}$ W/cm$^2$.}	\label{Fluid_vel}			
\end{figure}

The fluid velocity evolution is shown in Fig.~\ref{Fluid_vel} for all three IR laser intensities.
The fluid velocity of the unperturbed material is equal to zero and the velocity of perturbed material can be positive (moving down along the z-direction) or negative (moving in the opposite direction).
On the top of the colloidal crystal the fluid velocity is negative which is a sign of ablation.
On the shock wavefront the velocity is positive, and the liquid polystyrene is moving inside the colloidal crystal.
When the shock wave stops the fluid velocity becomes negative due to the reflection of the shock wave. The maximum fluid velocity is observed during the first 100~ps of the shock wave propagation (see Fig.~\ref{Fluid_vel}).
The maximum fluid velocity is on the order of 2~km/s and is summarized in Table~1 in the main text.

\subsection{Simulations using HELIOS code only}

We performed another set of the hydrodynamic simulations for three IR laser intensity using only HELIOS program (Set~2).
The simulation Set~2 was performed in order to compare PIConGPU coupled to HELIOS simulations and pure HELIOS simulations.

In the simulation Set~2 the experimental laser parameters were used.
The simulations were performed for three experimental laser intensities ($I_1$=$3.0 \cdot 10^{14}$~W/cm$^2$,  (b)~$I_2$=$4.8 \cdot 10^{14}$~W/cm$^2$, (c)~$I_3$=$6.3 \cdot 10^{14}$~W/cm$^2$) with 800~nm wavelength.
The FWHM of the laser pulse was 50~fs and the peak laser power was set at 50~fs after the beginning of the simulation.

Percentage of power reflected at the critical surface was obtained from the PIConGPU simulations.
The energy emitted by IR laser and the energy absorbed by electrons for three IR laser intensities  is summarized in Table~\ref{Absorption} and the absorption coefficient was around 10$\%$ for all three IR laser intensities.
Therefore the reflection coefficient in the HELIOS simulations was set to 90$\%$.
This value of the reflection coefficient is also confirmed by simple estimations.
In case of dense plasma the absorption coefficient derived from the Fresnel formulas can be written in the form $A=4\pi l_s /\lambda$, where $A$ is the absorption coefficient, $l_s$ is the skin depth and $\lambda$ is the IR laser wavelength~\cite{gamaly2002ablation}.
For the estimated skin depth of 10~nm the absorption coefficient is about 15~$\%$.

In the simulation Set~2, the two-temperature model was used, similar to the previous case.
In the two-temperature model both electrons and ions were assumed to have a room temperature in the initial state.
All other parameters of the simulation, except the laser parameters were the same as in the previous case (simulation Set~1 summarized in the Methods section in the main text).
The initial mass density of the colloidal crystal was similar to the previous simulation and had the same periodicity.
The quiet start temperature was set to 0.044~eV which is equal to the polystyrene melting temperature.

\begin{figure}[h]
	\includegraphics[width=1.0\linewidth]{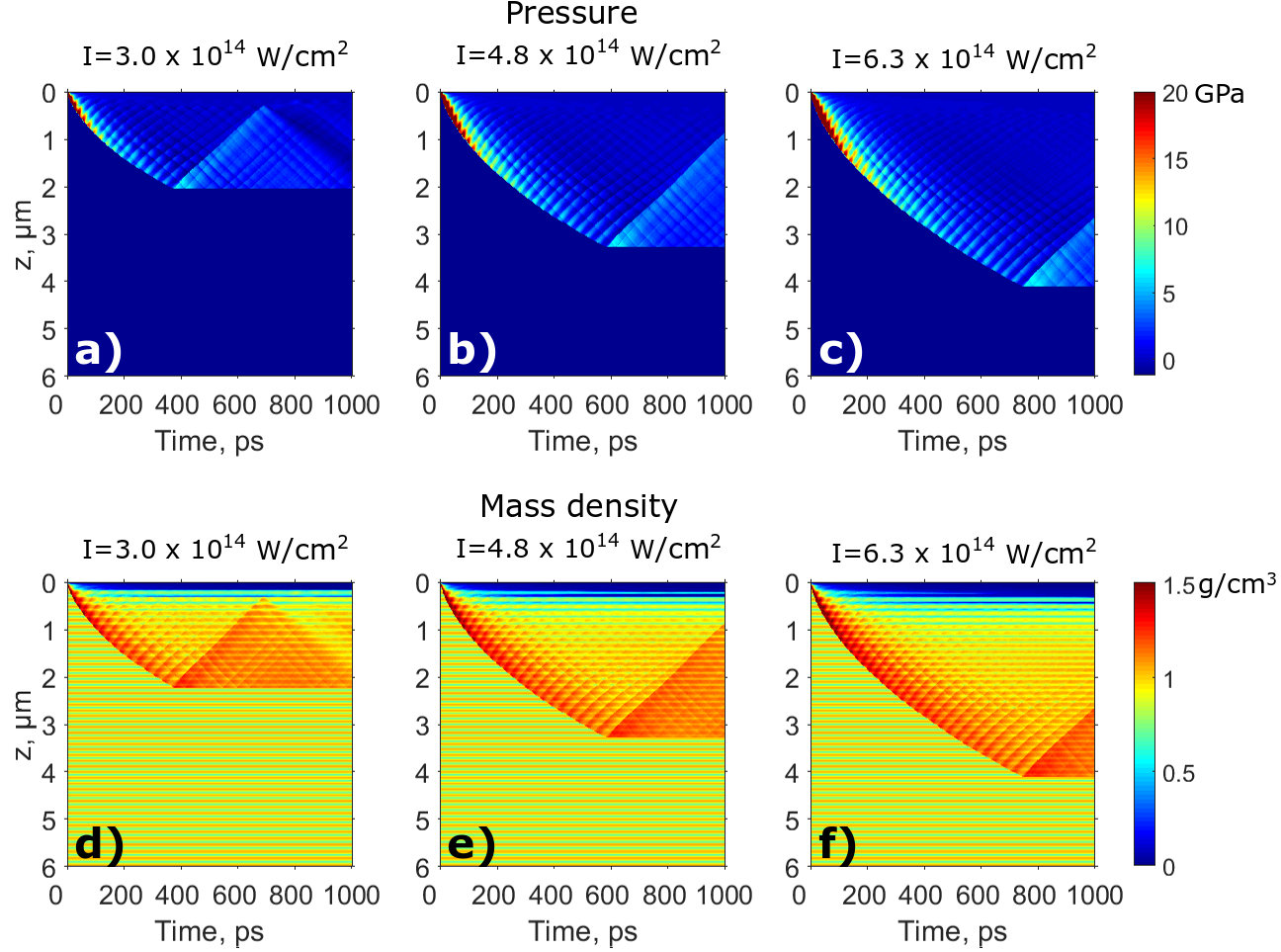}
	\caption{ Hydrodynamic simulations of the shock wave propagation inside the periodic colloidal crystal. The simulations were performed using  only the HELIOS code. Color plots show simulation results for the pressure  (a-c) and mass density (d-f) for three different IR laser intensities: (a,d)~$I_1$=$3.0 \cdot 10^{14}$~W/cm$^2$,  (b,e)~$I_2$=$4.8\cdot 10^{14}$~W/cm$^2$, (c-f)~$I_3$=$6.3 \cdot 10^{14}$~W/cm$^2$. }			
	\label{HELIOS_laser}			
\end{figure}

\begin{figure}[h]
	\includegraphics[width=0.85\linewidth]{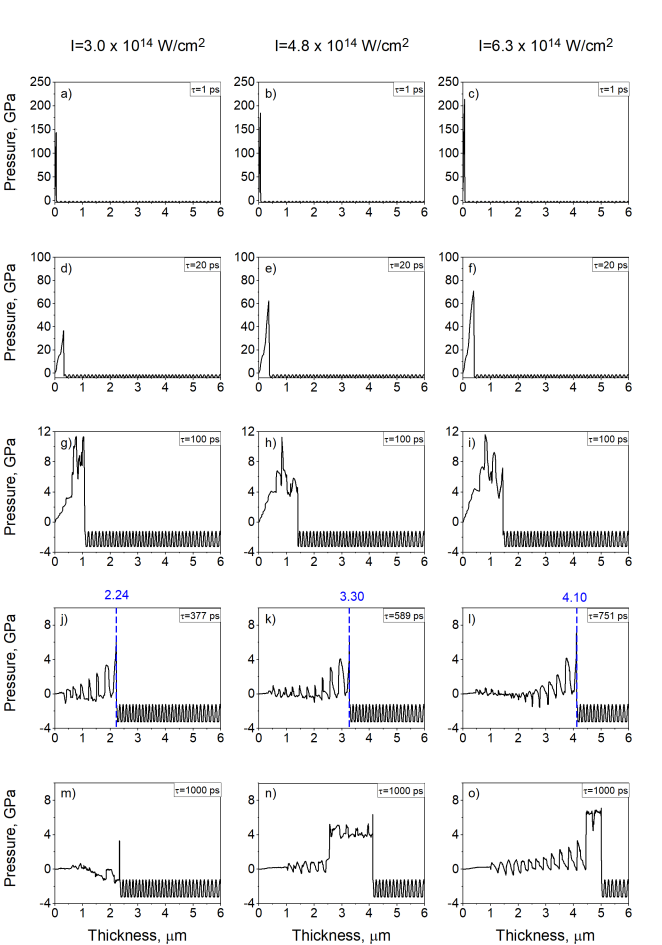}
	\caption{Pressure in the colloidal crystal from the hydrodynamic simulation.  The simulations were performed using  only the HELIOS code. Results are shown at 1~ps (a-c), 20 ps (d-f), 100~ps (g-i) after the stop of the shock wave (j-l) and at 1000~ps, the end of the simulation (m-o). The blue dashed line shows the shock wave stop depth (j-l). }			
	\label{Pressure_laser}			
\end{figure}

\begin{figure}[h]
	\includegraphics[width=0.85\linewidth]{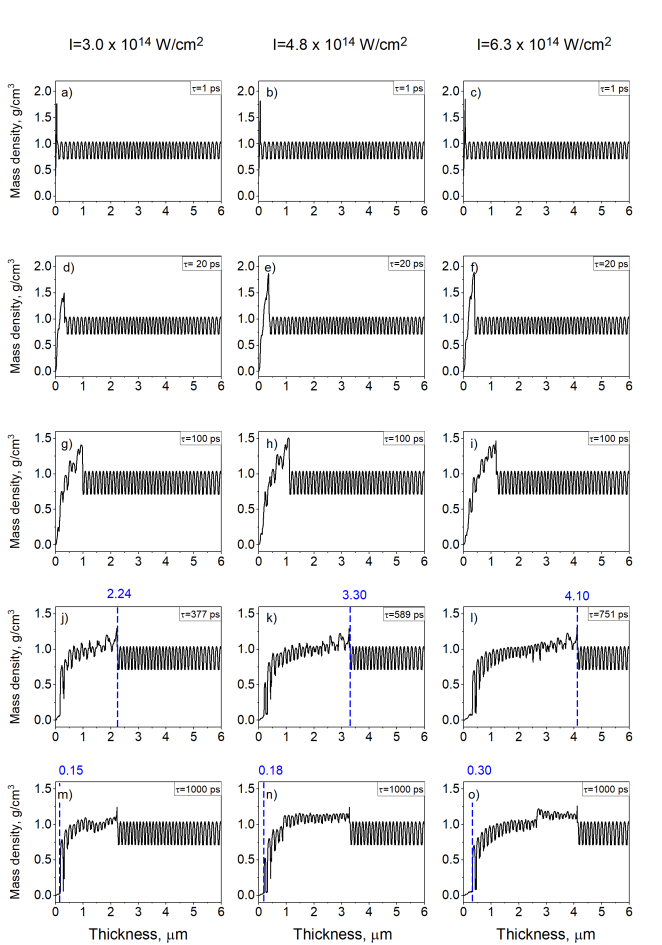}
	\caption{ Mass density of the periodic colloidal crystal from the hydrodynamic simulation. In this case only the HELIOS code was used. Results are shown at 1~ps (a-c), 20 ps (d-f), 100~ps (g-i) after the stop of the shock wave (j-l) and at 1000~ps, the end of the simulation (m-o). The blue dashed line shows the shock wave stop depth (j-l) and the ablation threshold (d-f). }			
	\label{Density_laser}			
\end{figure}

The pressure and mass density obtained from these set of simulations are shown in Figs.~\ref{HELIOS_laser},\ref{Pressure_laser},\ref{Density_laser}.
Two processes occur in the simulated colloidal crystal sample - ablation and shock propagation (the same processes were observed in simulation Set~1).
Ablation is well seen in Fig.~\ref{HELIOS_laser}(d-e) as the zero mass density on the top of the sample.
The top 1-2 layers are ablated during the first picoseconds and the ablation depth is summarized in Table~\ref{Table_laser}.
In the simulation Set~2 the ablation depth is slightly smaller than in the simulation Set~1 (see Fig.~5 in the main text and Fig.~\ref{Mass_1D}).
Such a difference may be caused by the different IR laser absorption mechanisms implemented in these programs.

The maximum shock wave pressure  achieved at the first picosecond of the simulation is on the order of 150-200~GPa (see Table~\ref{Table_laser} and Fig.~\ref{Pressure_laser}(a-c)).
At 20~ps the shock wavefront propagated through 0.5~$\mu$m of the colloidal crystal and destroyed the periodic structure of the sample.
In the simulation Set~1 the periodicity of the first 1~$\mu$m was damaged 20~picoseconds after the beginning of the laser pulse.
Therefore the simulation Set~2 is not explaining the fast drop of the diffracted intensity in the experimental results, while the simulation Set~1 the changes in the colloidal crystal structure can be attributed to the reduced diffracted intensity.

Further the shock wave is propagating through the colloidal crystal sample.
The shock wave stops at the different time and depth summarized in Table~\ref{Table_laser}.
It is worth to notice that compare to simulation Set~1, here the shock wave stops earlier and destroys less of the material.
The difference in the shock wave depth is on the order of 25$\%$ for two higher IR laser intensities and approximately 1$\%$ for the lower IR laser intensity.
Such a difference in the shock induced dynamics might be due to the different geometries (1D in HELIOS or 3D in PIConGPU).
In the 3D PIConGPU simulations the colloidal crystal structure was properly set, while in the HELIOS simulation only a 1D projection of the mass density was used.

\begin{table}[h]
	\setlength{\tabcolsep}{10pt}
	\caption{ The IR laser emitted and absorbed energy, and absorption coefficient estimated from PIConGPU simulations.  }
	\begin{tabular}{|l|c|c|c|}
		\hline
		Intensity, 10$^{14}$ W/cm$^2$  & 3.0 & 4.8 & 6.3 \\ \hline
		Laser emitted energy, 10$^{10}$ eV & 3.5 & 5.5 & 7.2 \\ \hline
		Absorbed energy, 10$^9$ eV & 3.5 & 5.7 & 7.0 \\ \hline
		Absorption coefficient, $\%$  & 10.0 & 9.6 & 9.7 \\ \hline
	\end{tabular}
	\label{Absorption}
\end{table}

\begin{table}[h]
	\setlength{\tabcolsep}{10pt}
	\caption{ Results of the simulation Set~2 (performed only with HELIOS code on the periodic colloidal crystal sample). The results are shown for three IR laser intensities. Ablation depth and shock wave time and depth were calculated from the results of the hydrodynamic simulations. }
	\begin{tabular}{|l|c|c|c|}
		\hline
		Intensity, 10$^{14}$ W/cm$^2$  & 3.0 & 4.8 & 6.3 \\ \hline
		Maximum shock wave pressure, GPa  & 143 & 184 & 214 \\ \hline
		Ablation depth, nm             & 150 & 180 & 300  \\ \hline
		Shock wave depth, $\mu$m        & 2.24 & 3.30 & 4.10   \\ \hline
		Simulated shock wave stop times, ps  & 377 & 589 & 751 \\ \hline
	\end{tabular}
	\label{Table_laser}
\end{table}

\subsection{HELIOS simulations for the bulk polystyrene sample}

\begin{figure}[h]
	\includegraphics[width=1.0\linewidth]{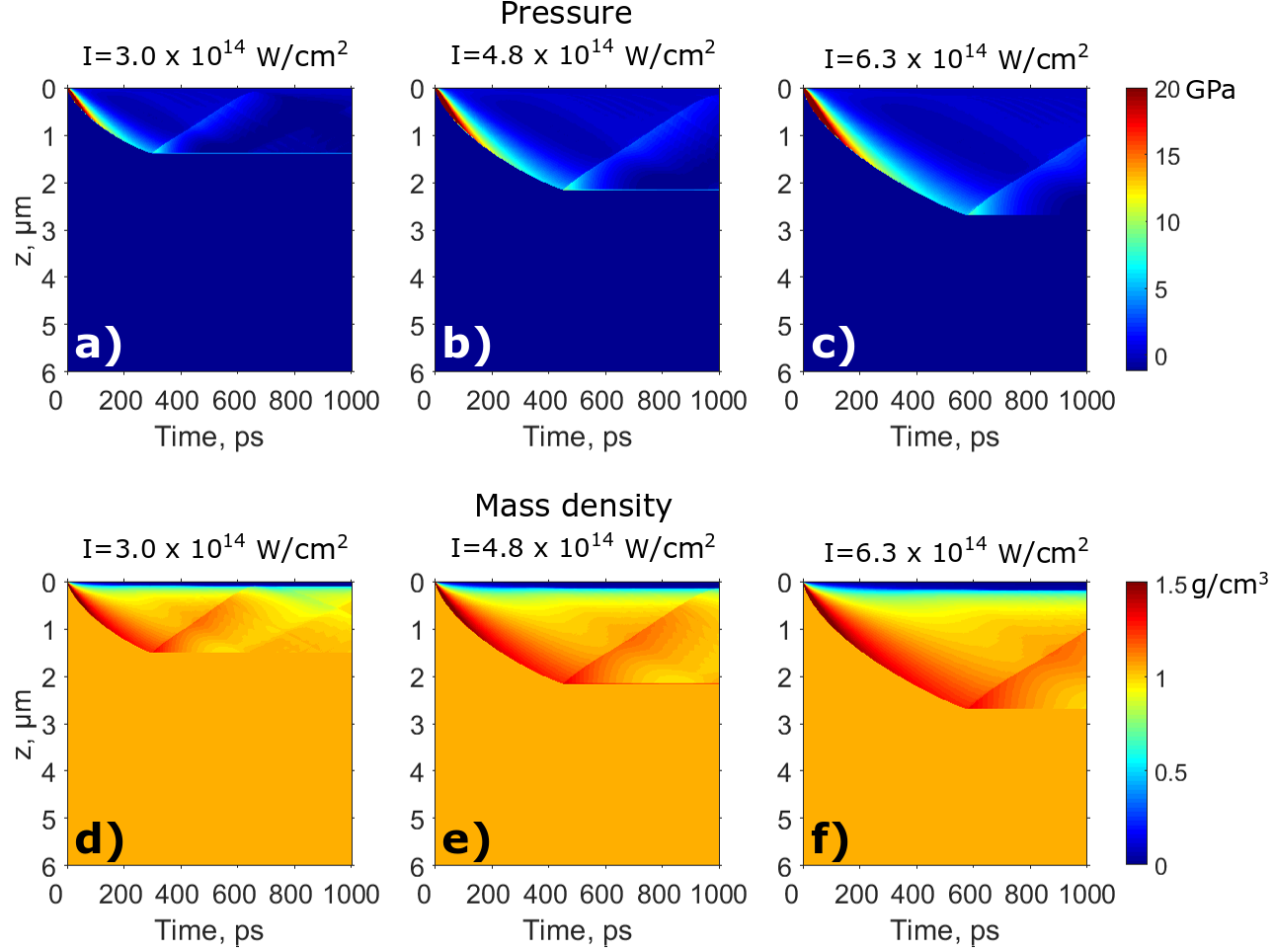}
	\caption{ Hydrodynamic simulations of the shock wave propagation inside the bulk polystyrene sample. In this case only the HELIOS code was used. Color plots show simulation results for the pressure  (a-c) and mass density (d-f) for three different IR laser intensities: (a,d)~$I_1$=$3.0 \cdot 10^{14}$ W/cm$^2$,  (b,e)~$I_2$=$4.8 \cdot 10^{14}$ W/cm$^2$, (c-f)~$I_3$=$6.3 \cdot 10^{14}$ W/cm$^2$. }			
	\label{HELIOS_bulk}			
\end{figure}

\begin{figure}[h]
	\includegraphics[width=0.85\linewidth]{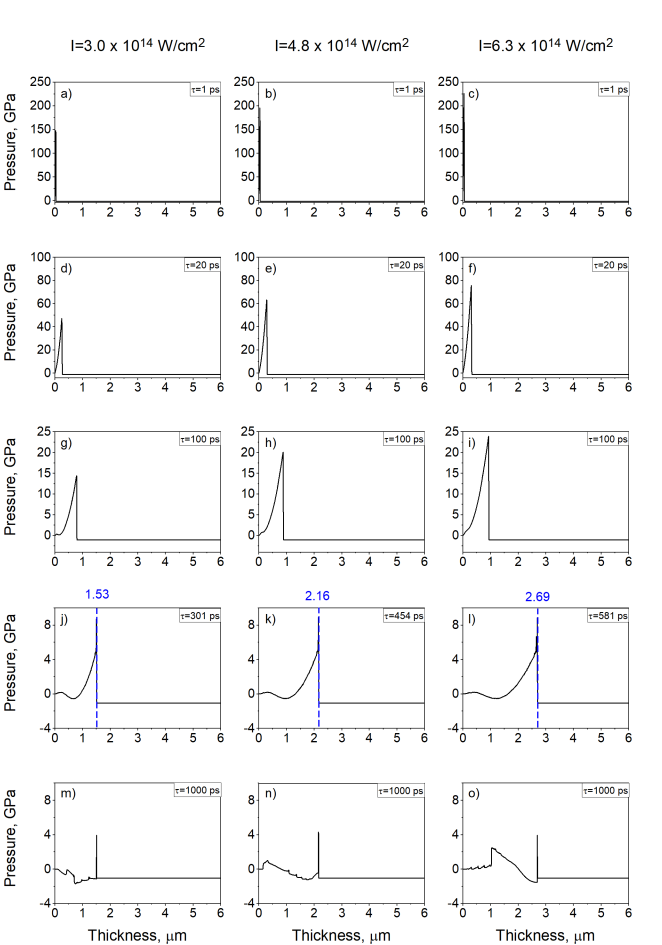}
	\caption{Pressure inside the of the bulk polystyrene from the hydrodynamic simulation. In this case only the HELIOS code was used. Results are shown at 1~ps (a-c), 20 ps (d-f), 100~ps (g-i) after the stop of the shock wave (j-l) and at 1000~ps, the end of the simulation (m-o). The blue dashed line shows the shock wave stop depth (j-l). }			
	\label{Pressure_bulk}			
\end{figure}

\begin{figure}[h]
	\includegraphics[width=0.85\linewidth]{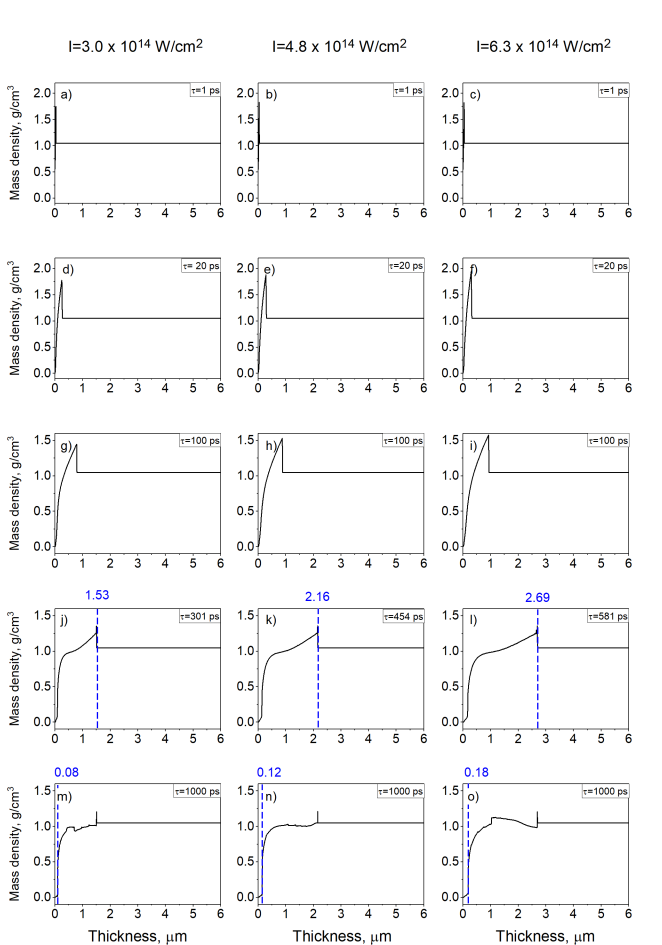}
	\caption{ Mass density of the bulk polystyrene from the hydrodynamic simulation. In this case only the HELIOS code was used. Results are shown at 1~ps (a-c), 20 ps (d-f), 100~ps (g-i) after the stop of the shock wave (j-l) and at 1000~ps, the end of the simulation (m-o). The blue dashed line shows the shock wave stop depth (j-l) and the ablation threshold (d-f). }			
	\label{Density_bulk}			
\end{figure}

The last set of the hydrodynamics simulations was performed using only HELIOS code with the bulk polystyrene sample (Set~3).
In this case all the simulations parameters were similar to the Set~2, except the initial polystyrene mass density.
The polystyrene mass density was set to be constant along z - direction (1.05~g/cm$^3$) considering bulk polystyrene.
The results of the simulation Set~3 are present in Fig.~\ref{HELIOS_bulk},~\ref{Pressure_bulk}, ~\ref{Density_bulk}.
The two main processes in the simulation Set~3 are ablation and the shock wave propagation, similar to simulation Set~1 and Set~2.

The snapshot of the simulation at 1~ps is shown in Fig.~\ref{HELIOS_bulk}(a-c) and ~\ref{Pressure_bulk}(a-c), and the initial constant pressure and density of the sample is visible.
The ablation of the material is already visible at 20~ps and it continues up to 100~ps (see Fig.~\ref{HELIOS_bulk},~\ref{Density_bulk}.
The ablation depth is on the order of 100-200~nm and is summarized in Table.~\ref{Table_bulk}.
In the simulation Set~3 the ablation depth was approximately twice smaller than for simulation Set~2.
The difference in the ablation depth can be explained by smaller average mass density in case of the simulation Set~2.

The maximum pressure is reached at the first picosecond of the simulation on the very top layer of the polystyrene sample (see Fig.~\ref{Pressure_bulk} and Table.~\ref{Table_bulk}).
The pressure--depth dependence had a periodic structure in the simulation Set~1 and Set~2 but in the simulation Set~3 it does not show any periodicity.
Therefore we can conclude that such pressure modulations were caused by the periodic structure of the colloidal crystal sample.

The peak pressure is decaying while the shock wave is propagating inside the sample (see Fig.~\ref{Pressure_bulk}), but the shape of the shock wavefront remains stable.
When the shock pressure is not sufficient to compress the material the shock wave stops and the shock wave stop time and depth are summarized in Table.~\ref{Table_bulk}.
For the simulation Set~3 the shock wave stop time are much smaller than in case of simulation Set~2.
The difference can be explained by the higher average mass density of the simulated sample.

\begin{table}[h]
	\setlength{\tabcolsep}{10pt}
	\caption{ Results of the simulation Set~3 (performed only with HELIOS code on the periodic colloidal crystal sample). The results are shown for three IR laser intensities. Ablation depth and shock wave time and depth were calculated from the results of the hydrodynamic simulations. }
	\begin{tabular}{|l|c|c|c|}
		\hline
		Intensity, 10$^{14}$ W/cm$^2$  & 3.0 & 4.8 & 6.3 \\ \hline
		Maximum shock wave pressure, GPa  & 148 & 196 & 225 \\ \hline
		Ablation depth, nm             & 80 & 120 & 180  \\ \hline
		Shock wave depth, $\mu$m        & 1.53 & 2.16 & 2.69   \\ \hline
		Simulated shock wave stop times, ps  & 301 & 454 & 581 \\ \hline
	\end{tabular}
	\label{Table_bulk}
\end{table}

\clearpage

\clearpage

\bibliography{References}

\begin{thebibliography}{10}
\expandafter\ifx\csname url\endcsname\relax
  \def\url#1{\texttt{#1}}\fi
\expandafter\ifx\csname urlprefix\endcsname\relax\def\urlprefix{URL }\fi
\providecommand{\bibinfo}[2]{#2}
\providecommand{\eprint}[2][]{\url{#2}}

\bibitem{briggs2017ultrafast}
\bibinfo{author}{Briggs, R.} \emph{et~al.}
\newblock \bibinfo{title}{Ultrafast x-ray diffraction studies of the phase
  transitions and equation of state of scandium shock compressed to 82 {GPa}}.
\newblock \emph{\bibinfo{journal}{Phys. Rev. Lett.}}
  \textbf{\bibinfo{volume}{118}}, \bibinfo{pages}{025501}
  (\bibinfo{year}{2017}).

\bibitem{stan2016liquid}
\bibinfo{author}{Stan, C.~A.} \emph{et~al.}
\newblock \bibinfo{title}{Liquid explosions induced by x-ray laser pulses}.
\newblock \emph{\bibinfo{journal}{Nature Physics}}
  \textbf{\bibinfo{volume}{12}}, \bibinfo{pages}{966} (\bibinfo{year}{2016}).

\bibitem{kraus2016nanosecond}
\bibinfo{author}{Kraus, D.} \emph{et~al.}
\newblock \bibinfo{title}{Nanosecond formation of diamond and lonsdaleite by
  shock compression of graphite}.
\newblock \emph{\bibinfo{journal}{Nat. Commun.}} \textbf{\bibinfo{volume}{7}},
  \bibinfo{pages}{10970} (\bibinfo{year}{2016}).

\bibitem{schropp2015imaging}
\bibinfo{author}{Schropp, A.} \emph{et~al.}
\newblock \bibinfo{title}{Imaging shock waves in diamond with both high
  temporal and spatial resolution at an {XFEL}}.
\newblock \emph{\bibinfo{journal}{Scientific Reports}}
  \textbf{\bibinfo{volume}{5}}, \bibinfo{pages}{11089} (\bibinfo{year}{2015}).

\bibitem{koenig2005progress}
\bibinfo{author}{Koenig, M.} \emph{et~al.}
\newblock \bibinfo{title}{Progress in the study of warm dense matter}.
\newblock \emph{\bibinfo{journal}{Plasma Phys. Controlled Fusion}}
  \textbf{\bibinfo{volume}{47}}, \bibinfo{pages}{B441} (\bibinfo{year}{2005}).

\bibitem{ross1981ice}
\bibinfo{author}{Ross, M.}
\newblock \bibinfo{title}{The ice layer in {Uranus} and {Neptune} diamonds in
  the sky?}
\newblock \emph{\bibinfo{journal}{Nature}} \textbf{\bibinfo{volume}{292}},
  \bibinfo{pages}{435--436} (\bibinfo{year}{1981}).

\bibitem{coppari2013experimental}
\bibinfo{author}{Coppari, F.} \emph{et~al.}
\newblock \bibinfo{title}{Experimental evidence for a phase transition in
  magnesium oxide at exoplanet pressures}.
\newblock \emph{\bibinfo{journal}{Nat. Geosci.}} \textbf{\bibinfo{volume}{6}},
  \bibinfo{pages}{926--929} (\bibinfo{year}{2013}).

\bibitem{benedetti1999dissociation}
\bibinfo{author}{Benedetti, L.~R.} \emph{et~al.}
\newblock \bibinfo{title}{Dissociation of {CH4} at high pressures and
  temperatures: diamond formation in giant planet interiors?}
\newblock \emph{\bibinfo{journal}{Science}} \textbf{\bibinfo{volume}{286}},
  \bibinfo{pages}{100--102} (\bibinfo{year}{1999}).

\bibitem{fletcher2015ultrabright}
\bibinfo{author}{Fletcher, L.} \emph{et~al.}
\newblock \bibinfo{title}{Ultrabright x-ray laser scattering for dynamic warm
  dense matter physics}.
\newblock \emph{\bibinfo{journal}{Nat. Photonics}}
  \textbf{\bibinfo{volume}{9}}, \bibinfo{pages}{274--279}
  (\bibinfo{year}{2015}).

\bibitem{drake2006high}
\bibinfo{author}{Drake, R.~P.}
\newblock \emph{\bibinfo{title}{High-energy-density physics: fundamentals,
  inertial fusion, and experimental astrophysics}}
  (\bibinfo{publisher}{Springer, Berlin}, \bibinfo{year}{2006}).

\bibitem{ciricosta2016measurements}
\bibinfo{author}{Ciricosta, O.} \emph{et~al.}
\newblock \bibinfo{title}{Measurements of continuum lowering in solid-density
  plasmas created from elements and compounds}.
\newblock \emph{\bibinfo{journal}{Nat. Commun.}} \textbf{\bibinfo{volume}{7}},
  \bibinfo{pages}{11713} (\bibinfo{year}{2016}).

\bibitem{kraus2017formation}
\bibinfo{author}{Kraus, D.} \emph{et~al.}
\newblock \bibinfo{title}{Formation of diamonds in laser-compressed
  hydrocarbons at planetary interior conditions}.
\newblock \emph{\bibinfo{journal}{Nat. Astron.}} \textbf{\bibinfo{volume}{1}},
  \bibinfo{pages}{606} (\bibinfo{year}{2017}).

\bibitem{kraus2016dynamic}
\bibinfo{author}{Kraus, R.} \emph{et~al.}
\newblock \bibinfo{title}{Dynamic compression of copper to over 450 {GPa}: A
  high-pressure standard}.
\newblock \emph{\bibinfo{journal}{Phys. Rev. B}} \textbf{\bibinfo{volume}{93}},
  \bibinfo{pages}{134105} (\bibinfo{year}{2016}).

\bibitem{ackermann2007operation}
\bibinfo{author}{Ackermann, W.~a.} \emph{et~al.}
\newblock \bibinfo{title}{Operation of a free-electron laser from the extreme
  ultraviolet to the water window}.
\newblock \emph{\bibinfo{journal}{Nat. Photonics}}
  \textbf{\bibinfo{volume}{1}}, \bibinfo{pages}{336} (\bibinfo{year}{2007}).

\bibitem{emma2010first}
\bibinfo{author}{Emma, P.} \emph{et~al.}
\newblock \bibinfo{title}{First lasing and operation of an
  {\aa}ngstrom-wavelength free-electron laser}.
\newblock \emph{\bibinfo{journal}{Nat. Photonics}}
  \textbf{\bibinfo{volume}{4}}, \bibinfo{pages}{641--647}
  (\bibinfo{year}{2010}).

\bibitem{ishikawa2012compact}
\bibinfo{author}{Ishikawa, T.} \emph{et~al.}
\newblock \bibinfo{title}{A compact x-ray free-electron laser emitting in the
  sub-{\aa}ngstr{\"o}m region}.
\newblock \emph{\bibinfo{journal}{Nat. Photonics}}
  \textbf{\bibinfo{volume}{6}}, \bibinfo{pages}{540} (\bibinfo{year}{2012}).

\bibitem{tschentscher2017photon}
\bibinfo{author}{Tschentscher, T.} \emph{et~al.}
\newblock \bibinfo{title}{Photon beam transport and scientific instruments at
  the {European} {XFEL}}.
\newblock \emph{\bibinfo{journal}{Applied Sciences}}
  \textbf{\bibinfo{volume}{7}}, \bibinfo{pages}{592} (\bibinfo{year}{2017}).

\bibitem{kluge2018observation}
\bibinfo{author}{Kluge, T.} \emph{et~al.}
\newblock \bibinfo{title}{Observation of ultrafast solid-density plasma
  dynamics using femtosecond x-ray pulses from a free-electron laser}.
\newblock \emph{\bibinfo{journal}{Phys. Rev. X}} \textbf{\bibinfo{volume}{8}},
  \bibinfo{pages}{031068} (\bibinfo{year}{2018}).

\bibitem{vinko2012creation}
\bibinfo{author}{Vinko, S.} \emph{et~al.}
\newblock \bibinfo{title}{Creation and diagnosis of a solid-density plasma with
  an x-ray free-electron laser}.
\newblock \emph{\bibinfo{journal}{Nature}} \textbf{\bibinfo{volume}{482}},
  \bibinfo{pages}{59--62} (\bibinfo{year}{2012}).

\bibitem{dronyak2012dynamics}
\bibinfo{author}{Dronyak, R.} \emph{et~al.}
\newblock \bibinfo{title}{Dynamics of colloidal crystals studied by pump-probe
  experiments at {FLASH}}.
\newblock \emph{\bibinfo{journal}{Phys. Rev. B}} \textbf{\bibinfo{volume}{86}},
  \bibinfo{pages}{064303} (\bibinfo{year}{2012}).

\bibitem{petukhov2003bragg}
\bibinfo{author}{Petukhov, A.}, \bibinfo{author}{Dolbnya, I.},
  \bibinfo{author}{Aarts, D.}, \bibinfo{author}{Vroege, G.} \&
  \bibinfo{author}{Lekkerkerker, H.}
\newblock \bibinfo{title}{Bragg rods and multiple x-ray scattering in
  random-stacking colloidal crystals}.
\newblock \emph{\bibinfo{journal}{Phys. Rev. Lett.}}
  \textbf{\bibinfo{volume}{90}}, \bibinfo{pages}{028304}
  (\bibinfo{year}{2003}).

\bibitem{sulyanova2015structural}
\bibinfo{author}{Sulyanova, E.~A.} \emph{et~al.}
\newblock \bibinfo{title}{Structural evolution of colloidal crystal films in
  the process of melting revealed by bragg peak analysis}.
\newblock \emph{\bibinfo{journal}{Langmuir}} \textbf{\bibinfo{volume}{31}},
  \bibinfo{pages}{5274--5283} (\bibinfo{year}{2015}).

\bibitem{shabalin2016revealing}
\bibinfo{author}{Shabalin, A.} \emph{et~al.}
\newblock \bibinfo{title}{Revealing three-dimensional structure of an
  individual colloidal crystal grain by coherent x-ray diffractive imaging}.
\newblock \emph{\bibinfo{journal}{Phys. Rev. Lett.}}
  \textbf{\bibinfo{volume}{117}}, \bibinfo{pages}{138002}
  (\bibinfo{year}{2016}).

\bibitem{gulden2010coherent}
\bibinfo{author}{Gulden, J.} \emph{et~al.}
\newblock \bibinfo{title}{Coherent x-ray imaging of defects in colloidal
  crystals}.
\newblock \emph{\bibinfo{journal}{Phys. Rev. B}} \textbf{\bibinfo{volume}{81}},
  \bibinfo{pages}{224105} (\bibinfo{year}{2010}).

\bibitem{bosak2010high}
\bibinfo{author}{Bosak, A.}, \bibinfo{author}{Snigireva, I.},
  \bibinfo{author}{Napolskii, K.~S.} \& \bibinfo{author}{Snigirev, A.}
\newblock \bibinfo{title}{High-resolution transmission x-ray microscopy: A new
  tool for mesoscopic materials}.
\newblock \emph{\bibinfo{journal}{Adv. Mater.}} \textbf{\bibinfo{volume}{22}},
  \bibinfo{pages}{3256--3259} (\bibinfo{year}{2010}).

\bibitem{van2011scanning}
\bibinfo{author}{van Schooneveld, M.~M.} \emph{et~al.}
\newblock \bibinfo{title}{Scanning transmission x-ray microscopy as a novel
  tool to probe colloidal and photonic crystals}.
\newblock \emph{\bibinfo{journal}{Small}} \textbf{\bibinfo{volume}{7}},
  \bibinfo{pages}{804--811} (\bibinfo{year}{2011}).

\bibitem{lehmann2017laser}
\bibinfo{author}{Lehmann, G.} \& \bibinfo{author}{Spatschek, K.}
\newblock \bibinfo{title}{Laser-driven plasma photonic crystals for high-power
  lasers}.
\newblock \emph{\bibinfo{journal}{‎Phys. Plasmas}}
  \textbf{\bibinfo{volume}{24}}, \bibinfo{pages}{056701}
  (\bibinfo{year}{2017}).

\bibitem{sakai2007properties}
\bibinfo{author}{Sakai, O.} \& \bibinfo{author}{Tachibana, K.}
\newblock \bibinfo{title}{Properties of electromagnetic wave propagation
  emerging in 2-d periodic plasma structures}.
\newblock \emph{\bibinfo{journal}{IEEE Transactions on Plasma Science}}
  \textbf{\bibinfo{volume}{35}}, \bibinfo{pages}{1267--1273}
  (\bibinfo{year}{2007}).

\bibitem{gildenburg2019grating}
\bibinfo{author}{Gildenburg, V.} \& \bibinfo{author}{Pavlichenko, I.}
\newblock \bibinfo{title}{Grating-like nanostructures formed by the focused fs
  laser pulse in the volume of transparent dielectric}.
\newblock \emph{\bibinfo{journal}{Optics letters}}
  \textbf{\bibinfo{volume}{44}}, \bibinfo{pages}{2534--2537}
  (\bibinfo{year}{2019}).

\bibitem{kuo2007enhancement}
\bibinfo{author}{Kuo, C.-C.} \emph{et~al.}
\newblock \bibinfo{title}{Enhancement of relativistic harmonic generation by an
  optically preformed periodic plasma waveguide}.
\newblock \emph{\bibinfo{journal}{Phys. Rev. Lett.}}
  \textbf{\bibinfo{volume}{98}}, \bibinfo{pages}{033901}
  (\bibinfo{year}{2007}).

\bibitem{monchoce2014optically}
\bibinfo{author}{Monchoc{\'e}, S.} \emph{et~al.}
\newblock \bibinfo{title}{Optically controlled solid-density transient plasma
  gratings}.
\newblock \emph{\bibinfo{journal}{Phys. Rev. Lett.}}
  \textbf{\bibinfo{volume}{112}}, \bibinfo{pages}{145008}
  (\bibinfo{year}{2014}).

\bibitem{leblanc2016ptychographic}
\bibinfo{author}{Leblanc, A.}, \bibinfo{author}{Monchoc{\'e}, S.},
  \bibinfo{author}{Bourassin-Bouchet, C.}, \bibinfo{author}{Kahaly, S.} \&
  \bibinfo{author}{Qu{\'e}r{\'e}, F.}
\newblock \bibinfo{title}{Ptychographic measurements of ultrahigh-intensity
  laser--plasma interactions}.
\newblock \emph{\bibinfo{journal}{Nature Physics}}
  \textbf{\bibinfo{volume}{12}}, \bibinfo{pages}{301} (\bibinfo{year}{2016}).

\bibitem{leblanc2017spatial}
\bibinfo{author}{Leblanc, A.} \emph{et~al.}
\newblock \bibinfo{title}{Spatial properties of high-order harmonic beams from
  plasma mirrors: a ptychographic study}.
\newblock \emph{\bibinfo{journal}{Phys. Rev. Lett.}}
  \textbf{\bibinfo{volume}{119}}, \bibinfo{pages}{155001}
  (\bibinfo{year}{2017}).

\bibitem{ganeev2014third}
\bibinfo{author}{Ganeev, R.}, \bibinfo{author}{Boltaev, G.} \&
  \bibinfo{author}{Usmanov, T.}
\newblock \bibinfo{title}{Third and fourth harmonics generation in
  laser-induced periodic plasmas}.
\newblock \emph{\bibinfo{journal}{Optics Communications}}
  \textbf{\bibinfo{volume}{324}}, \bibinfo{pages}{114--119}
  (\bibinfo{year}{2014}).

\bibitem{rudenko2017femtosecond}
\bibinfo{author}{Rudenko, A.} \emph{et~al.}
\newblock \bibinfo{title}{Femtosecond response of polyatomic molecules to
  ultra-intense hard x-rays}.
\newblock \emph{\bibinfo{journal}{Nature}} \textbf{\bibinfo{volume}{546}},
  \bibinfo{pages}{129} (\bibinfo{year}{2017}).

\bibitem{abbey2016x}
\bibinfo{author}{Abbey, B.} \emph{et~al.}
\newblock \bibinfo{title}{X-ray laser--induced electron dynamics observed by
  femtosecond diffraction from nanocrystals of buckminsterfullerene}.
\newblock \emph{\bibinfo{journal}{Sci. Adv.}} \textbf{\bibinfo{volume}{2}},
  \bibinfo{pages}{e1601186} (\bibinfo{year}{2016}).

\bibitem{kraus2018high}
\bibinfo{author}{Kraus, D.} \emph{et~al.}
\newblock \bibinfo{title}{High-pressure chemistry of hydrocarbons relevant to
  planetary interiors and inertial confinement fusion}.
\newblock \emph{\bibinfo{journal}{‎Phys. Plasmas}}
  \textbf{\bibinfo{volume}{25}}, \bibinfo{pages}{056313}
  (\bibinfo{year}{2018}).

\bibitem{helled2010interior}
\bibinfo{author}{Helled, R.}, \bibinfo{author}{Anderson, J.~D.},
  \bibinfo{author}{Podolak, M.} \& \bibinfo{author}{Schubert, G.}
\newblock \bibinfo{title}{Interior models of {Uranus} and {Neptune}}.
\newblock \emph{\bibinfo{journal}{Astrophys. J.}}
  \textbf{\bibinfo{volume}{726}}, \bibinfo{pages}{15} (\bibinfo{year}{2010}).

\bibitem{madhusudhan2012possible}
\bibinfo{author}{Madhusudhan, N.}, \bibinfo{author}{Lee, K.~K.} \&
  \bibinfo{author}{Mousis, O.}
\newblock \bibinfo{title}{A possible carbon-rich interior in super-earth 55
  cancri e}.
\newblock \emph{\bibinfo{journal}{The Astrophysical Journal Letters}}
  \textbf{\bibinfo{volume}{759}}, \bibinfo{pages}{L40} (\bibinfo{year}{2012}).

\bibitem{dufour2007white}
\bibinfo{author}{Dufour, P.}, \bibinfo{author}{Liebert, J.},
  \bibinfo{author}{Fontaine, G.} \& \bibinfo{author}{Behara, N.}
\newblock \bibinfo{title}{White dwarf stars with carbon atmospheres}.
\newblock \emph{\bibinfo{journal}{Nature}} \textbf{\bibinfo{volume}{450}},
  \bibinfo{pages}{522} (\bibinfo{year}{2007}).

\bibitem{kraus2013probing}
\bibinfo{author}{Kraus, D.} \emph{et~al.}
\newblock \bibinfo{title}{Probing the complex ion structure in liquid carbon at
  100 {GPa}}.
\newblock \emph{\bibinfo{journal}{Phys. Rev. Lett.}}
  \textbf{\bibinfo{volume}{111}}, \bibinfo{pages}{255501}
  (\bibinfo{year}{2013}).

\bibitem{chollet2015x}
\bibinfo{author}{Chollet, M.} \emph{et~al.}
\newblock \bibinfo{title}{The x-ray pump--probe instrument at the linac
  coherent light source}.
\newblock \emph{\bibinfo{journal}{J. Synchrotron Radiat.}}
  \textbf{\bibinfo{volume}{22}}, \bibinfo{pages}{503--507}
  (\bibinfo{year}{2015}).

\bibitem{mukharamova2017probing}
\bibinfo{author}{Mukharamova, N.} \emph{et~al.}
\newblock \bibinfo{title}{Probing dynamics in colloidal crystals with
  pump-probe experiments at {LCLS}: Methodology and analysis}.
\newblock \emph{\bibinfo{journal}{Applied Sciences}}
  \textbf{\bibinfo{volume}{7}}, \bibinfo{pages}{519} (\bibinfo{year}{2017}).

\bibitem{meijer2015colloidal}
\bibinfo{author}{Meijer, J.-M.}
\newblock \emph{\bibinfo{title}{Colloidal Crystals of Spheres and Cubes in Real
  and Reciprocal Space}} (\bibinfo{publisher}{Springer}, \bibinfo{year}{2015}).

\bibitem{keldysh1965ionization}
\bibinfo{author}{Keldysh, L.}
\newblock \bibinfo{title}{Ionization in the field of a strong electromagnetic
  wave}.
\newblock \emph{\bibinfo{journal}{Sov. Phys. JETP}}
  \textbf{\bibinfo{volume}{20}}, \bibinfo{pages}{1307--1314}
  (\bibinfo{year}{1965}).

\bibitem{reiss2014tunnelling}
\bibinfo{author}{Reiss, H.}
\newblock \bibinfo{title}{The tunnelling model of laser-induced ionization and
  its failure at low frequencies}.
\newblock \emph{\bibinfo{journal}{Journal of Physics B: Atomic, Molecular and
  Optical Physics}} \textbf{\bibinfo{volume}{47}}, \bibinfo{pages}{204006}
  (\bibinfo{year}{2014}).

\bibitem{reiss2008limits}
\bibinfo{author}{Reiss, H.}
\newblock \bibinfo{title}{Limits on tunneling theories of strong-field
  ionization}.
\newblock \emph{\bibinfo{journal}{Phys. Rev. Lett.}}
  \textbf{\bibinfo{volume}{101}}, \bibinfo{pages}{043002}
  (\bibinfo{year}{2008}).

\bibitem{kittel2005introduction}
\bibinfo{author}{Kittel, C.}
\newblock \emph{\bibinfo{title}{Introduction to solid state physics}}
  (\bibinfo{publisher}{Wiley, New York}, \bibinfo{year}{2005}).

\bibitem{pitaevskii2012physical}
\bibinfo{author}{Pitaevskii, L.} \& \bibinfo{author}{Lifshitz, E.}
\newblock \emph{\bibinfo{title}{Physical kinetics}}, vol.~\bibinfo{volume}{10}
  (\bibinfo{publisher}{Butterworth-Heinemann}, \bibinfo{year}{2012}).

\bibitem{krainov2002cluster}
\bibinfo{author}{Krainov, V.~P.} \& \bibinfo{author}{Smirnov, M.~B.}
\newblock \bibinfo{title}{Cluster beams in the super-intense femtosecond laser
  pulse}.
\newblock \emph{\bibinfo{journal}{Phys. Rep.}} \textbf{\bibinfo{volume}{370}},
  \bibinfo{pages}{237--331} (\bibinfo{year}{2002}).

\bibitem{gibbon1996short}
\bibinfo{author}{Gibbon, P.} \& \bibinfo{author}{F{\"o}rster, E.}
\newblock \bibinfo{title}{Short-pulse laser-plasma interactions}.
\newblock \emph{\bibinfo{journal}{Plasma Phys. Controlled Fusion}}
  \textbf{\bibinfo{volume}{38}}, \bibinfo{pages}{769} (\bibinfo{year}{1996}).

\bibitem{PIConGPU2013}
\bibinfo{author}{Bussmann, M.} \emph{et~al.}
\newblock \bibinfo{title}{Radiative signatures of the relativistic
  kelvin-helmholtz instability}.
\newblock In \emph{\bibinfo{booktitle}{Proceedings of the International
  Conference on High Performance Computing, Networking, Storage and Analysis}},
  SC '13, \bibinfo{pages}{5:1--5:12} (\bibinfo{year}{2013}).

\bibitem{burau2010picongpu}
\bibinfo{author}{Burau, H.} \emph{et~al.}
\newblock \bibinfo{title}{{PIConGPU}: A fully relativistic particle-in-cell
  code for a {GPU} cluster}.
\newblock \emph{\bibinfo{journal}{IEEE Transactions on Plasma Science}}
  \textbf{\bibinfo{volume}{38}}, \bibinfo{pages}{2831--2839}
  (\bibinfo{year}{2010}).

\bibitem{delone1998tunneling}
\bibinfo{author}{Delone, N.~B.} \& \bibinfo{author}{Krainov, V.~P.}
\newblock \bibinfo{title}{Tunneling and barrier-suppression ionization of atoms
  and ions in a laser radiation field}.
\newblock \emph{\bibinfo{journal}{Physics-Uspekhi}}
  \textbf{\bibinfo{volume}{41}}, \bibinfo{pages}{469--485}
  (\bibinfo{year}{1998}).

\bibitem{more1985pressure}
\bibinfo{author}{More, R.}
\newblock \bibinfo{title}{Pressure ionization, resonances, and the continuity
  of bound and free states}.
\newblock \emph{\bibinfo{journal}{Advances in atomic and molecular physics}}
  \textbf{\bibinfo{volume}{21}}, \bibinfo{pages}{305--356}
  (\bibinfo{year}{1985}).

\bibitem{macfarlane2006helios}
\bibinfo{author}{MacFarlane, J.}, \bibinfo{author}{Golovkin, I.} \&
  \bibinfo{author}{Woodruff, P.}
\newblock \bibinfo{title}{{HELIOS-CR}--a {1-D} radiation-magnetohydrodynamics
  code with inline atomic kinetics modeling}.
\newblock \emph{\bibinfo{journal}{Journal of Quantitative Spectroscopy and
  Radiative Transfer}} \textbf{\bibinfo{volume}{99}}, \bibinfo{pages}{381--397}
  (\bibinfo{year}{2006}).

\bibitem{suriano2011femtosecond}
\bibinfo{author}{Suriano, R.} \emph{et~al.}
\newblock \bibinfo{title}{Femtosecond laser ablation of polymeric substrates
  for the fabrication of microfluidic channels}.
\newblock \emph{\bibinfo{journal}{Applied Surface Science}}
  \textbf{\bibinfo{volume}{257}}, \bibinfo{pages}{6243--6250}
  (\bibinfo{year}{2011}).

\bibitem{marsh1980lasl}
\bibinfo{author}{Marsh, S.~P.}
\newblock \emph{\bibinfo{title}{{LASL} Shock Hugoniot Data}},
  vol.~\bibinfo{volume}{5} (\bibinfo{publisher}{University of California Press,
  Berkeley}, \bibinfo{year}{1980}).

\bibitem{gamaly2013physics}
\bibinfo{author}{Gamaly, E.~G.} \& \bibinfo{author}{Rode, A.~V.}
\newblock \bibinfo{title}{Physics of ultra-short laser interaction with matter:
  From phonon excitation to ultimate transformations}.
\newblock \emph{\bibinfo{journal}{Progress in Quantum Electronics}}
  \textbf{\bibinfo{volume}{37}}, \bibinfo{pages}{215--323}
  (\bibinfo{year}{2013}).

\bibitem{hart2012cspad}
\bibinfo{author}{Hart, P.} \emph{et~al.}
\newblock \bibinfo{title}{The {CSPAD} megapixel x-ray camera at{ LCLS}}.
\newblock In \emph{\bibinfo{booktitle}{Proc. SPIE}}, vol.
  \bibinfo{volume}{8504}, \bibinfo{pages}{85040C} (\bibinfo{year}{2012}).

\bibitem{courant1928partial}
\bibinfo{author}{Courant, R.}, \bibinfo{author}{Friedrichs, K.} \&
  \bibinfo{author}{Lewy, H.}
\newblock \bibinfo{title}{On the partial difference equations op mathematical
  physics}.
\newblock \emph{\bibinfo{journal}{Mathematische Annalen}}
  \textbf{\bibinfo{volume}{100}}, \bibinfo{pages}{32--74}
  (\bibinfo{year}{1928}).

\bibitem{yee1966numerical}
\bibinfo{author}{Yee, K.}
\newblock \bibinfo{title}{Numerical solution of initial boundary value problems
  involving maxwell's equations in isotropic media}.
\newblock \emph{\bibinfo{journal}{IEEE Transactions on antennas and
  propagation}} \textbf{\bibinfo{volume}{14}}, \bibinfo{pages}{302--307}
  (\bibinfo{year}{1966}).

\bibitem{huebl_axel_2015_33624}
\bibinfo{author}{Huebl, A.} \emph{et~al.}
\newblock \bibinfo{title}{{openPMD 1.0.0: A meta data standard for particle and
  mesh based data.}} (\bibinfo{year}{2015}).

\bibitem{gamaly2002ablation}
\bibinfo{author}{Gamaly, E.~G.}, \bibinfo{author}{Rode, A.~V.},
  \bibinfo{author}{Luther-Davies, B.} \& \bibinfo{author}{Tikhonchuk, V.~T.}
\newblock \bibinfo{title}{Ablation of solids by femtosecond lasers: Ablation
  mechanism and ablation thresholds for metals and dielectrics}.
\newblock \emph{\bibinfo{journal}{Physics of plasmas}}
  \textbf{\bibinfo{volume}{9}}, \bibinfo{pages}{949--957}
  (\bibinfo{year}{2002}).

\end{thebibliography}

\end{document}